\documentstyle[12pt,aaspp4]{article}

\def\msun{${\rm M}_\odot$ }
\def\la{\mathrel{\hbox{\rlap{\hbox{\lower4pt\hbox{$\sim$}}}\hbox{$<$}}}}
\def\ga{\mathrel{\hbox{\rlap{\hbox{\lower4pt\hbox{$\sim$}}}\hbox{$>$}}}}

\begin{document}


\title{Evidence for Harmonic Relationships in the High Frequency QPOs
of XTE~J1550--564 and GRO~J1655--40}


\author{Ronald A. Remillard and Michael P. Muno}
\affil{Center for Space Research, MIT, Cambridge, MA 02139 \\ rr@space.mit.edu; muno@space.mit.edu}

\and

\author{Jeffrey E. McClintock}
\affil{Harvard-Smithsonian Center for Astrophysics, 60
Garden St. MS-3, \\ Cambridge, MA 02138; jem@cfa.harvard.edu}

\and

\author{Jerome A. Orosz}
\affil{Astronomical Institute, Utrecht University, Postbus 80000, 3508 TA Utrecht, \\ The Netherlands; J.A.Orosz@astro.uu.nl}

\begin{abstract}

We continue to investigate the X-ray timing and spectral properties of
the black hole binary XTE~J1550--564 during its 1998-1999 outburst. By
grouping observations according to the type of low-frequency
quasiperiodic oscillation (LFQPO) identified in a previous paper, we
show evidence that two high-frequency QPOs (HFQPOs) occur
simultaneously near 184 and 276 Hz.  In each case, we can model the
QPO profiles while assuming that the central frequencies are related
by a 3:2 ratio. In one group, there is also evidence of a broad
feature at the fundamental frequency of 92 Hz.  We also investigate
the April 2000 outburst of the source and confirm the suggestion of
Miller et al. that a 270 Hz QPO is accompanied by a second feature
near 180 Hz. We conclude that the majority of the 28 individual HFQPO
detections in XTE~J1550--564 are best interpreted as harmonics of a
fundamental oscillation near 92 Hz. The most significant exceptions to
this pattern are the 143 Hz QPO detected on 1998 October 15 and the 65
Hz QPO reported by Kalemci et al. for 2000 May 20. We next investigate
all of the X-ray energy spectra and QPO results for individual
observations of XTE~J1550--564, and we find a systematic increase in
the strength of the power-law component as the stronger of the two
HFQPOs shifts from 276 to 184 Hz. A strikingly similar result is seen
in the spectra of GRO~J1655--40 when the stronger HFQPO shifts from
450 to 300 Hz. The fundamental HFQPO frequencies for the two X-ray
sources scale as $M^{-1}$, which is consistent with the hypotheses
that these HFQPOs represent some kind of oscillation rooted in general
relativity (GR) and that the two black holes have similar values of
the dimensionless spin parameter. We discuss physical mechanisms that
may explain these HFQPOs. A resonance between orbital and radial
coordinate frequencies is one possibility suggested by Abramowicz \&
Kluzniak.  For XTE~J1550-564, this would imply moderate values for the
dimensionless spin parameter ($0.1 < a_* < 0.6$), with similar results
for GRO~J1655-40. However, there remain large uncertainties as to
whether this model can produce the HFQPOs and their variations in
frequency.  We also discuss models for ``diskoseismic'' oscillations.
In this case, the concept that the inner disk behaves as a resonance
cavity in GR has certain attractions for explaining HFQPOs, but
integral harmonics are not predicted for the three types of
diskoseismic modes.

\end{abstract}


\section{Introduction}

XTE~J1550--564 is a recurrent X-ray nova and a microquasar that has
been extensively studied at X-ray and optical frequencies.  A major
outburst during 1998-1999 revealed complex evolution in both its
spectral properties and variability characteristics (\cite{sob00b};
\cite{hom01}; \cite{rem02}). It is one of five black hole candidates
that exhibit transient high-frequency QPOs (HFQPOs), while its
stronger and more common low-frequency QPOs (LFQPOs) are seen to vary
in frequency in response to changes in the spectrum of the accretion
disk. Additional X-ray outbursts with weaker maxima and shorter
duration were observed in 2000 April to June (\cite{tom01b};
\cite{mil01}), in 2001 February - March (\cite{tom01a}), and again in
2002 January (\cite{swa02}).  The accumulated investment of almost 400
RXTE pointed observations (2-200 keV) provides a unique archive of
high-quality monitoring data that span 4 decades in X-ray luminosity,
with several excursions through each of the canonical emission states
of black hole binaries. These resources motivate continuing efforts to
understand the structure of the accretion flow at various levels of
X-ray luminosity and to search for effects of general relativity (GR)
in the innermost regions of the accretion disk.

Optical photometry of XTE~J1550--564 during outbursts revealed the
binary period of 1.54 days and the perplexing weakness of an optical
response to X-ray flares (\cite{jai01}). Optical spectroscopy and
photometry during accretion quiescence has recently established that
the system is a dynamical black hole binary, with a black hole mass
of $10.0 \pm 1.5$ \msun (\cite{oro02}). The companion
star is a late type subgiant (G8IV to K4III), and the binary
inclination angle is constrained to the range $72^\circ \pm
5^\circ$ (\cite{oro02}).

Radio observations revealed a relativistic jet associated with the
large X-ray flare of 1998 September 19 (\cite{han01}). The radio
source then decayed away while the X-ray source remained in the ``very
high'' state.  During the X-ray ``low-hard'' state seen in the
outburst of 2000, weaker radio flux was detected with an inverted
spectrum, suggesting the presence of a steady jet (\cite{cor01}).

In a previous paper (\cite{rem02}; hereafter RSMM02), we used 209 RXTE
observations of XTE~J1550--564 during its 1998-1999 outburst to
investigate the relationships between HFQPOs, LFQPOs and the spectral
characteristics reported by Sobczak et al. (2000b). \nocite{sob00b} It
was found that complex LFQPOs are better organized in their
correlations with both HFQPOs and the accretion disk flux and
temperature when we distinguished three types (``A'', ``B'', and
``C'') on the basis of phase lags and coherence values measured at the
central QPO frequency. The LFQPO types (but not their frequencies) are
correlated with the frequencies of HFQPOs that were detected on 20
occasions. Type A LFQPOs (5--10 Hz, broad profiles, lag in soft
X-rays) are associated with narrow HFQPOs near 276 Hz.  Type B LFQPOs
(5--7 Hz, narrow profiles, lag in hard X-rays) correspond to broader
HFQPOs near 184 Hz. The common type ``C'' LFQPOs (variable
frequencies, narrow profiles, strong amplitude and harmonics, and
small phase lags) coincide on rare occasions with HFQPOs at
lower frequencies (100--169 Hz).

There are now five black hole candidates that exhibit transient
HFQPOs, and for three of them there is evidence that two HFQPOs can
occur simultaneously.  HFQPO pairs have been seen in GRO~J1655--40
(300, 450 Hz; \cite{str01a}; \cite{rem99b}), GRS~1915+105 (40, 67 Hz;
\cite{str01b}), and also probably during the 2000 outburst of
XTE~J1550--564 (\cite{mil01}). In the first two cases, Strohmayer
(2001a; b) investigated combinations of the azimuthal and radial
coordinate frequencies in GR to explain the pair of HFQPOs, noting
that other explanations are possible. For the case of GRO~J1655--40,
it was further claimed (\cite{str01a}) that the 450 Hz QPO suggests
that the black hole must have substantial spin (i.e.  dimensionless
spin parameter, $a_* > 0.15$. This conclusion uses the optically
determined mass of the black hole (\cite{sha99}; \cite{gre01}), along
with the assumption that the orbital rotation frequency (in GR) at the
innermost stable circular orbit around the black hole (\cite{sha83})
is the highest frequency that can be seen in the X-rays emission.

An alternative interpretation for pairs of HFQPOs utilizes
``diskoseismology'', which considers adiabatic perturbations in a
relativistic accretion disk (\cite{wag99}; \cite{kat01}). If the
HFQPOs in GRO~J1655--40 and GRS~1915+105 represent fundamental
$g-$mode and $c-$mode diskoseismic oscillations, then substantially
higher values of the spin parameter are derived ($a_* \sim 0.9 and
\sim 0.7$, respectively; \cite{wag01}). For the case of GRO~J1655--40,
yet another interpretation was offered by Abramowicz \& Kluzniak
(2001) \nocite{abr01}, who hypothesize that there is enhanced X-ray
emission at the radius in the accretion disk where there is a
resonances between the Keplerian and radial coordinate
frequencies. This idea is motivated by the 3:2 integral ratio seen in
the frequencies of HFQPOs from that source. When considering all of
the HFQPO observations, it is possible that the results may require
more than one physical model.

As a followup to RSMM02, we address four questions related to QPO
behavior in XTE~J1550--564. (1) If we average the power density
spectra (PDS) during the 1998-1999 outburst for groups defined by the
LFQPO type, is there any evidence for a pair of HFQPOs? (2)
Can we confirm the suggestion of Miller et al. (2001) \nocite{mil01}
that HFQPOs near 180 Hz and 270 Hz occur simultaneously during the
outburst of 2000? (3) Are the pairs of HFQPOs in XTE~J1550--564 and
GRO~J1655--40 related precisely by a 3:2 ratio? (4) Are there X-ray
spectral properties that distinguish the observations when the upper
or lower HFQPO is stronger in each of these two sources?

\section{Observations and Data Analysis}

We continue the analysis of XTE~J1550--564 using RXTE observations
reported in previous publications. For the 209 observations of the
1998-1999 outburst, spectral parameters were tabulated by Sobczak et
al. (2000b). \nocite{sob00b} QPO properties are given in
RSMM02. Timing results for 19 observations (XTE program P50134) during
the outburst of 2000 April-May are reported by Miller et al. (2001).
\nocite{mil01} We supplement these with analyses of contemporaneous
exposures under RXTE programs P50135 (37 observations) and P50137 (11
observations), which are now publicly available.

We elaborate on our methods for conducting error analysis for features
in the PDS, since the significance of QPOs is an important topic in
$\S 3$. As noted in RSMM02, the PDS ($P_{\nu}$ vs. $\nu$) are computed
for 256 s data segments and then averaged for each observation.  We
first compute the PDS using the normalization of Leahy et al. (1983)
\nocite {lea83}.  For each frequency bin in the discrete Fourier
transform ($\Delta \nu = 2^{-8}$ Hz), we compute the uncertainty as
the larger of either the statistical error ($2/N^{0.5}$) or the
observed standard deviation of the mean power, where there are $N$ data
segments within the observation. We then subtract the
Poisson-corrected dead time, as described in Morgan, Remillard, \&
Greiner (1997), \nocite{mor97} using the average count rates (per PCU)
for good events and for very large events (which have a longer dead
time) during a given observation.  The PDS are then renormalized by
the mean source count rate (i.e. above the non-source background), so that
the final PDS has units of (rms deviation / mean)$^2$ Hz$^{-1}$. We
logarithmically rebin the frequencies using intervals that are
successively larger by a factor of 1.04, propagating the uncertainty
within each frequency interval. When we average the PDS for groups of
observations, we compute the weighted mean (using $\sigma^{-2}$) and
its uncertainty for each frequency interval.

The PDS are searched for LFQPOs (0.5 to 30 Hz) and HFQPOs (30 to 1000
Hz) separately, using a sliding frequency window that typically spans
0.2 to 5.0 $\nu$. Within the window, the power continuum ($P_{cont}$)
is modeled with a second order polynomial in log $\nu$, since the local
power continuum usually resembles a power-law function with a slight
amount of broad curvature. The QPO profiles are presumed to be
Lorentzian functions, and they are generally distinguished from broad
peaks in the power continuum by a coherence parameter, $Q = \nu / FWHM
\ga 2$. We use $\chi^2$ minimization to obtain the best fit for the
QPO profile and the local power continuum.  The uncertainty in the
power continuum is estimated via $\chi^2$ analysis near the best-fit
value, and a similar treatment is used to estimate the uncertainties
in the QPO central frequency ($\nu_0$) and the $FWHM$.

If the inclusion of a QPO feature is statistically warranted, and if
the best-fit value of $\chi_{\nu}^2$ is acceptable, then we evaluate
the significance of the feature with a conservative and empirical
approach, as follows. Within the central frequency range of the QPO,
$\nu_0 \pm FWHM$, we integrate $P_{\nu} - P_{cont}$ and then divide
the result ($S$) by its statistical uncertainty ($\sigma_S$), where
the calculation of $\sigma_S$ considers, in quadrature, the
uncertainties in the power density measurements as well as the
continuum fit.  Usually, $\sigma_S$ is dominated by the measurement
uncertainty in the power density bins, since there are many bins that
constrain the continuum fit.  However, QPO searches may be susceptible
to systematic errors in the power continuum model (especially for
broad QPOs), which are difficult to ascertain.  For this reason, and
given our practical experience searching for QPOs in many black hole
candidates, we judge the threshold for a high level of confidence to
be near a significance $S/\sigma_S \ga 4$ for QPO searches using this
method.

When QPO detections are significant, we use the best-fit values for
the Lorentzian peak and $FWHM$ to compute the total integrated power
($P$) in the PDS feature, and the rms amplitude of the QPO is then $a
= P^{0.5}$.  We can then use the QPO significance as a measure of the
uncertainty in $P$, leading to an estimate of the amplitude
uncertainty, $\sigma_a / a = 0.5 \sigma_S / S$. We note that the
factor of 0.5 was inadvertently neglected in reporting $\sigma_a$
values (overly large) in Table 1 of Remillard et al. (1999a)
\nocite{rem99a}.

We note that global PDS modeling has been performed successfully for
sources in hard X-ray states using multiple Lorentzians (\cite
{now00}; \cite {pot02}).  The results yield broad profiles ($Q <
0.5$), which distinguish these features as broad power peaks rather
than QPOs. We note, however, that the multiple-Lorentzian model
generally does not work well for PDS associated with the ``high'' or
``very high'' states where the power continuum roughly resembles a
power-law function over several decades in frequency. This is the case
for most of the observations of XTE~J1550--564 that exhibit HFQPOs,
and therefore a more localized fit for the power continuum is a
practical necessity.

\section{Results}

We show the results for investigations of the four questions raised in
the $\S 1$.  All of the power density spectra are displayed in units
of log($\nu \times P_{\nu}$) vs. log($\nu$), thereby making it easier
to see peaks in the power density and to evaluate the relative
strengths of features at widely different frequencies (\cite{psa99};
\cite{now00}). We note, however, that all of the QPO fits are computed
in the ($P_{\nu}, \nu$)-plane.

\subsection{PDS of XTE~J1550--564 during 1998-1999, Averaged by LFQPO Type}

Using Table 1 of RSMM02, we computed the average power spectra of
XTE~J1550--564 (6 to 30 keV) for groups of observations that show LFQPO
types A (10 cases), B (9 cases), and C (44 of 46 cases that have data
in this energy band).  Observations classified ``A?'' were included in
the A group, but the five ``C\'~'' cases that followed the 7.7 Crab
flare were not included in the C group.  The results are shown in
Fig.~\ref{fig:pds1550}. We have utilized the energy range 6 to 30 keV
for this analysis, because this energy range optimizes the detections of
the various HFQPOs seen in XTE~J1550--564.  Analogous investigations
at 2 to 30 keV or 13 to 30 keV produce similar conclusions with
weaker statistics.

In RSMM02, the the broad type A LFQPOs (5--10 Hz with phase lags in
soft X-rays) were associated with a narrow HFQPO near 280 Hz. In
Fig.~\ref{fig:pds1550} this HFQPO is the strongest feature in the
average PDS (top panel), while there is also evidence of a weak
feature near 185 Hz which is not detected during any individual
observation. The narrow type B LFQPOs (5--7 Hz, strong harmonics, hard
lags) are associated with a broader HFQPO near 185 Hz. In
Fig.~\ref{fig:pds1550}, the average B-type PDS clearly shows the
LFQPO, its harmonics at $0.5 \nu_0$ and $2 \nu_0$, and the strong peak
at 185 Hz. There is also complex structure and possibly additional
HFQPOs on both sides of the 185 Hz feature. Type C LFQPOs ($\nu_0$
varying from 0.08 to 7 Hz, weak phase lags) are smeared out in the
average C-type PDS, which is dominated by a continuum that may be
deconvolved into three or more broad power peaks. For the C-type PDS,
the integrated rms fractional variability (0.1 to 1000 Hz) is 0.21 at
6-30 keV, while the total power for types A and B is only 0.04 to 0.08
at 6-30 keV. Finally, the 4th panel of Fig.~\ref{fig:pds1550} shows
the PDS for 1998 September 19, when XTE~J1550--564 was in the midst of
a flare to a peak intensity of 7 Crab (see
e.g. \cite{sob00b}). Significant QPOs are seen at 4.9, 13.2, and 183
Hz on that occasion.

To quantify the significance of the weaker high-frequency features in
the type A and B average PDS, we modeled the data in the range 40-1200
Hz according to the procedures described in $\S 2$. In each of these
fits, the central frequencies are forced to have ratios of 2:3 or
1:2:3, and their positions are adjusted with one free parameter. The
results are shown in Fig.~\ref{fig:mdl1550}.  The top two panels show
QPO fits for the type A and B groups; the PDS data are the same as
those plotted in Fig.~\ref{fig:pds1550}. The best fit models for the
QPO profiles are shown with dark lines, the power continuum is shown
with a dashed line, and the central QPO frequencies in the model are
shown with vertical ticks above the data.  We note that our dead time
correction model chronically leaves residual power at the level of
$P_{\nu} \sim 10^{-6}$ Hz$^{-1}$ (see \cite{rev00} for further
discussions of this effect). Thus, with the PDS displayed in log($\nu
\times P_{\nu}$) units, the power continuum (dashed line) shows some
upward curvature while smoothly connecting the source power at 50 Hz
to the residual dead time effects at 1 kHz.

All of the QPO parameters derived from these profile models are given
in Table 1. The significance of the 187 Hz QPO in the average type A
PDS is near 4 $\sigma$, as is the shallow QPO found at 92 Hz in the
type B PDS. We note that these are near the level deemed to be
reliable for general QPO searches. We conclude that there is sound
evidence for harmonic relationships in the HFQPOs in XTE~J1550--564,
but we also acknowledge the need for confirmation of these results for
this source and for other black hole binaries as well.  The type B PDS
seems to exhibit the highest harmonic content. We note the hints of
weak features seen at the 4th and 5th harmonics (2.4 $\sigma$ and 2.0
$\sigma$, respectively) as shown with arrows in
Fig.~\ref{fig:mdl1550}.
 
\subsection{HFQPOs in XTE~J1550--564 during the Outburst of 2000}

We also examine data from the second outburst of XTE~J1550--564
that began around 2000 April 4. 
Miller et al. (2001) \nocite{mil01} reported six detections of a QPO
near 270 Hz out of a total of 12 RXTE observations that were made between
2000 April 30 and May 9. This time interval occurs just after
the time of maximum luminosity (April 29), as seen with the RXTE All
Sky Monitor.  In the average PDS for these 6 observations, QPOs were
found at $268 \pm 3$ Hz (7.8 $\sigma$) and $188 \pm 3$ Hz (3.5
$\sigma$), suggesting that the features appear simultaneously
(\cite{mil01}).

Here, we add two details to the findings of Miller et al.  (2001), who
reported strong HFQPO detections by selecting PCA data above 6
keV. First, we find one additional HFQPO detection at $267 \pm 8$ Hz
($Q = 4.9$) on 2000 May 5. The significance is highest ($4.5 \sigma$)
for data in the energy range 2 to 30 keV. Second, for the five
remaining observations during 2000 April 30 to May 9, we find that
their average PDS contains a significant QPO near 270 Hz. We therefore
combine all 12 observations (6 to 30 keV) to further test the
significance of a second QPO near 180 Hz and to help judge the
acceptability of a model with forced harmonic relationships in the
central frequencies. The results (Fig.~\ref{fig:mdl1550} and Table 1)
support the conclusion by Miller et al. that the pair of QPOs occur
simultaneously, and the significance of the second QPO near 180 Hz is
increased to 5.0 $\sigma$ by averaging together all 12 observations.

The central frequencies (independently fit) derived by Miller et
al. (2001) deviate from a ratio of 3:2 by 2.2 $\sigma$. For the 12
observations in the same time interval (bottom panel of
Fig.~\ref{fig:mdl1550}), the harmonic fit is acceptable. Comparing the
QPO profile model with the data, one can see that the peak at lower
frequency might move to slightly higher frequency if it were treated
as a free parameter. Since the six QPO detections (\cite{mil01})
decrease in central frequency with time (276 to 249 Hz), small
deviations from a 3:2 frequency ratio could arise from changes in the
relative strength of the two features during the sampling interval.
We conclude that the QPO properties during the 2000 outburst of
XTE~J1550--564 are amenable to interpretation via a harmonic
relationship, but alternative possibilities cannot be excluded.

\subsection{Summary of HFQPOs in XTE~J1550--564}

There are now 28 HFQPO detections reported for individual observations
of XTE~J1550--564 (\cite{hom01}; RSMM02; \cite{mil01}; \cite{kal01};
$\S 3.2$). Their central frequencies vary substantially, but the
majority tend to cluster near 184 or 276 Hz. This is evident in the
histogram of HFQPO frequencies shown in Fig.~\ref{fig:hist} (top
panel). In making the histogram, we varied the binning 
intervals to maintain a width of $\pm$5 \% relative to the
central frequency.

None of the individual observations yield detections of a pair of
HFQPOs, yet when groups of observations are averaged, either by the
LFQPO type (1998-1999) or by time interval (2000), we find three
HFQPOs organized as harmonics near 92, 184, and 276 Hz. These results
suggest that the HFQPO harmonics appear simultaneously, but that only
the dominant feature is is detectable in the individual observations.

We next assess the extent to which any individual HFQPO detections
occur well outside of this harmonic scheme.  In most cases (17 of 28),
the allowed range of frequencies ($\nu \pm 1 \sigma$) lies within 3\%
of a harmonic value. At the other extreme, 3 of 28 cases deviate from
a harmonic value by an amount greater than 12\%. One of these is the
65 Hz QPO reported by Kalemci et al. (2001) \nocite{kal01} for 2000
May 20. The two others were observed on 1998 October 15 (with type C
LFQPOs), and their average PDS is shown in Fig.~\ref{fig:oct15}. The
QPO (5.0 $\sigma$) is found at $143 \pm 8$ Hz, which is closer to the
geometric mean of the harmonics than it is to either 92 or 184
Hz. Even though the statistical quality of the data is limited, it is
important to note that the harmonic scheme may represent only a subset
of the behavior patterns of HFQPOs in XTE~J1550--564.

\subsection{X-ray Spectra and HFQPOs in XTE~J1550--564}

We now focus attention on the spectral properties that may distinguish
which harmonic frequency is present in each observation that exhibits
an HFQPO. XTE~J1550--564 displays a typical X-ray spectrum for a black
hole binary, with a thermal component from the accretion disk and a
hard power-law attributed to inverse Compton scattering by electrons
of unknown origin. Sobczak et al. (2000b) deconvolved these spectral
components and reported parameters for the 209 RXTE observations
obtained during the 1998-1999 outburst.  We performed the same
analysis for the 43 RXTE observations obtained during 2000 in order to
compare the behavior of the source during the two outbursts.

It has been shown that the exercise of plotting the integrated disk
flux versus power-law flux in black hole binaries is very useful in
tracking the flow of energy between the two components through
different X-ray states, and also in relating QPO parameters to
spectral parameters (\cite{mun99}; \cite{sob00a}). Using the spectral
parameters of Sobczak et al. (2000b), \nocite{sob00b} we show the flux
measures in the two spectral components during the 1998-1999 outburst
in Fig.~\ref{fig:sp1550}. As discussed in RSMM02, one can integrate
the spectral components either over the PCA energy range with highest
sensitivity (2-25 keV) or over bolometric limits, in which case the
majority of the disk flux is extrapolated below 2 keV. Both options
are displayed in Fig.~\ref{fig:sp1550}.  For the bolometric estimates,
the lower limit on the integration of the power-law component is
reduced to 1 keV, which affects the results significantly when the
power law spectrum is steep.

We note that the bolometric disk flux is $2.16 \times 10^{-11} N_{dbb}
T^4$ erg cm$^{-2}$ s$^{-1}$, where $N_{dbb}$ is the multi-temperature
disk normalization: $N_{dbb} = R_{in} cos \theta / d^2 $, with the
inner disk radius, $R_{in}$, expressed in km and the distance, $d$, in
units of 10 kpc. Assuming a disk inclination $\theta = 72^\circ$ and
$d = 6$ kpc (\cite{oro02}), then the bolometric disk luminosity would
be $1.3 \times 10^{39}$ erg s$^{-1}$ (or 1.0 $L_{Edd}$ for a 10 \msun
black hole) for a value of $17.9 \times 10^{-8}$ erg cm$^{-2}$
s$^{-1}$ in the left panel (horizontal axis) in
Fig.~\ref{fig:sp1550}. For the power-law component, the luminosity
depends only on $d$, and the corresponding value (vertical axis) is
$29.2 \times 10^{-8}$ erg cm$^{-2}$ s$^{-1}$.

In Fig.~\ref{fig:sp1550}, we use the symbol color to denote the
following QPO conditions: HFQPO detections (blue), LFQPOs only (green
``x''), and no QPOs (red ``x'').  In addition, the shape of the blue
symbols distinguishes the HFQPOs near 92 Hz (triangle), 184 Hz (filled
square), and 276 Hz (star). The results for the pair of observations
on 1998 October 15 are displayed with blue circles.  There are clear
patterns in these results, and the conclusions do not depend on the
integration limits used to calculate the flux (compare left and right
panels).  There is a fairly well defined horizontal branch (red
``x''s) in which the disk dominates the spectrum and no QPOs are seen,
just as in the case of GRS~1915+105 (\cite{mun99}).  The large
majority of vertical excursions from this branch display some kind of
QPO activity. These spectra with enhanced power-law flux occur over a
wide range of disk flux, in contrast with the vertical branch at very
low disk flux in the case of GRS~1915+105 (\cite{mun99}).

The spectral diagrams in Fig.~\ref{fig:sp1550} also distinguish the
HFQPOs and the selection of different harmonic frequencies at
different times. There is a systematic difference in magnitude of
the power-law flux between the points associated with the 276 Hz QPO
(stars) versus the 184 Hz QPO (solid squares). The X-ray spectra
associated with the 276 Hz QPO lie systematically closer to the
horizontal branch in which the spectrum is dominated by the
accretion disk.

We present the same diagrams for the much weaker outburst of 2000 in
Fig.~\ref{fig:sp1550b}. This outburst samples different X-ray states
and reaches a maximum luminosity that is a factor of 10 below the
levels seen during 1998-1999.  Here, all of the HFQPO detections
associated with the 276 Hz feature (blue stars) occur at approximately
the same disk and power-law fluxes as their counterparts in
Fig.~\ref{fig:sp1550}, and they again lie close to the disk-dominant
branch (red x's) plotted in Fig. 4.  The observations during the time
interval of 2000 April 30 - May 9 correspond to the 12 red and blue
points in the lower-right quadrant of the right panel in
Fig.~\ref{fig:sp1550b}. We have shown that the average PDS for these
observations exhibits a strong QPO at 270 Hz accompanied by a $5
\sigma$ detection at 180 Hz (Fig.~\ref{fig:mdl1550} and Table 1). We
conclude that these spectral and temporal results for the 2000
outburst support the idea that HFQPOs near 184 Hz and 276 Hz represent
a temporal signature of accretion that is inherent to this black hole
binary system.

Finally, we note that XTE~J1550-564 additionally displays a vertical
branch (green x's) during the 2000 outburst (Fig.~\ref{fig:sp1550b})
in which the power law component varies while there is little
contribution from the disk. At these times the PDS show only LFQPOs
and strong continuum power. These observations are further associated
with a steady jet seen in the radio band (\cite{cor01}). A similar
X-ray track associated with a flat radio spectrum has been seen in
GRS~1915+105 (\cite{mun99}).

\subsection{Summary of HFQPOs in GRO~J1655--40}

HFQPOs near 300 Hz at photon energies of 2 to 30 keV were reported for
6 observations of GRO~J1655--40 by Remillard et al. (1999b),
\nocite{rem99b} (significance $\ga 3 \sigma$). The
examination of RXTE pointings under programs 10261 and 20187 (which
were not available for that paper) yield an
additional strong detection (6.0 $\sigma$) at $320 \pm 11$ Hz (with $Q
= 4.6$) for the observation on 1996 November 7.  There is no detection
of a QPO near 450 Hz in this observation.

Strohmayer (2001) reported detections of 5 HFQPOs near 450 Hz for PDS
computed using data from photon energies in the range 13 to 27
keV. Again, the analysis of the data from these other programs yields
an additional, weak detection (3.4 $\sigma$ at 13 to 27 keV) at $447
\pm 5$ Hz (with $Q = 8.9$) for the observation on 1996 June 20.  There
is no 300 Hz QPO detection during that observation.

These results raise the total to 10 observations in which one or more
HFQPOs are detected in GRO~J1655--40: 4 show only the 300 Hz QPO, 3
show only the 450 Hz QPO, and 3 show both. For the three observations
that exhibit both HFQPOs (\cite{str01a}), the independent profile fits
(using different energy bands) yield central frequencies of $295 \pm
4$ Hz (2-30 keV) and $440 \pm 5$ Hz (13-30 keV), respectively. These
results are consistent with a 3:2 ratio, as noted in $\S 1$, while one
feature is usually much stronger than the other, as seen in the
behavior of XTE~J1550--564. In the next section, we examine whether
the X-ray spectra evolve in parallel with the relative strength of the
HFQPO features in the case of GRO~J1655--40.

\subsection{X-ray Spectra and HFQPOs in GRO~J1655--40}

We use the 52 spectral observations reported by Sobczak et al. (1999)
\nocite{sob99}, supplemented by similar analyses for RXTE programs
10261 (six spectra binned in 1-day intervals), 20187 (1996 November
7), and 20402 (the final three observations, not reported
previously). In addition, since there is so much exposure time during
1996 May 9-11, we choose to sample the data in 7 time intervals,
rather than in 3 daily bins.  This yields a net of 66 spectral
observations of GRO~J1655--40.  We display the results of the spectral
decomposition in Fig.~\ref{fig:sp1655}.  The procedures are the same
as those conducted for XTE~J1550--564, except that we use an
inclination angle of $70^\circ$ and a distance of 3.2 kpc for
GRO~J1655--40 (\cite{gre01} and references therein).  In this case,
the disk luminosity would be $7.9 \times 10^{38}$ erg s$^{-1}$ (or 1.0
$L_{Edd}$ for a 6.3 \msun black hole) for a bolometric disk flux of
$45 \times 10^{-8}$ erg cm$^{-2}$ s$^{-1}$ in the left panel
(horizontal axis) of Fig.~\ref{fig:sp1655}. For the power-law
component, the same luminosity corresponds to $65 \times 10^{-8}$ erg
cm$^{-2}$ s$^{-1}$ in the same Figure panel (vertical axis).

For GRO~J1655--40, the relationship between HFQPO properties and the
energy division between the spectral components resembles
the results for XTE~J1550--564 in several important ways. The
observations with HFQPOs (blue symbols) lie above the horizontal
branch (red ``x''s) where the disk dominates the spectrum and no QPOs
are seen.  Moreover, with increasing power-law luminosity we see the
following progression: first are the observations where we detect only
the 450 Hz QPO (open stars), next are cases in which both HFQPOs are
detected simultaneously (filled stars), finally there are those in
which only the 300 Hz QPO is seen (solid squares). This is the same
pattern seen for XTE~J1550--564, where the strongest modulation
shifts from 276 to 184 Hz as the power-law flux increases
(cf. Fig.~\ref{fig:sp1550}). This similarity suggests that the
physical process that produces HFQPOs in these two sources is
identical.

\section{Discussion}

The 3:2:1 frequency ratio with a fundamental frequency near 92 Hz
seems to account for most of the X-ray HFQPOs detected in 28
individual observations of XTE~J1550--564. The features seen in the
average PDS for three well-defined groups of observations support this
scheme and further suggest that the HFQPO harmonics may coexist.
Moreover, all of the 13 HFQPO detections (in 10 observations) in
GRO~J1655--40 conform to a 3:2 frequency ratio with an implied
fundamental at 150 Hz, including the three occasions in which both the
300 Hz and 450 Hz QPOs were detected (\cite{str01a}). These results
provide solid evidence for harmonic relationships between the HFQPOs
in black hole binaries; nevertheless we acknowledge the need to
confirm this conclusion in these and other X-ray sources, since many
of the detections are near or below the statistical threshold for high
levels of confidence.

We have found that the concurrent evolution of the HFQPOs and the
X-ray spectra is very similar for the black hole binaries
XTE~J1550--564 and GRO~J1655--40.  The primary result is an increase
in the luminosity of the X-ray power-law component as the HFQPOs
exhibit a shift from the 3rd harmonic to the 2nd harmonic.  We also
note that HFQPOs are generally not detected along the spectral tracks
in which either the accretion disk or the power-law component strongly
dominates the spectrum (i.e. the horizontal, red tracks in
Figs. ~\ref{fig:sp1550} and ~\ref{fig:sp1655} and the vertical, green
track in Fig.~\ref{fig:sp1550b}). The energy decomposition diagrams do
not show perfect segregation of points relative to the HFQPO
properties. However, we cannot expect perfect organization, given the
variations in statistical sensitivity and the faintness of the HFQPOs
(rms amplitudes of 0.5\% to 5\% of the mean count rate in the selected
energy band).

The deduced fundamental frequencies for the HFQPOs in XTE~J1550--564
(92 Hz) and GRO~J1655--40 (150 Hz) differ by a factor of $1.63 \pm
0.06$. On the other hand, the inverse ratio of black hole masses is
$1.59 \pm 0.27$ (\cite{oro02} ; \cite{gre01}), which suggests that the
fundamental frequencies scale as $M^{-1}$. This result is generally
consistent with the known mechanisms related to disk
oscillations in the strong-field regime of general relativity, as long
as the values of the spin parameter ($a_*$) are similar for these two
black holes. These results illustrate both the quantitative value of
HFQPO detections as a means of probing the physical properties of
black holes, and also the need to continue efforts to independently
measure black hole masses via dynamical optical studies.

If these HFQPOs are indeed harmonically related, then we must attempt
to specify a physical model that can account for the value of the
fundamental frequency, an emission mechanism that produces HFQPOs
especially at the 2nd and 3rd harmonics, and an explanation for the
spectral evolution associated with the harmonic order. The results for
XTE~J1550--564 further suggest the need to accommodate occasional
shifts in the frequency system (e.g. $\sim 10$ \% shifts occur
$\sim 15$ \% of the time and perhaps $\sim 30$ \% shifts occur on rare
occasions). With regard to the largest shifts, it is also possible that
some HFQPOs involve a different physical mechanism, as suggested for
the HFQPO pair at 40 and 67 Hz (i.e., not harmonically related) in
GRS~1915+105 (\cite{str01b}). Finally, if the oscillations originate
in the inner accretion disk, there is a need to understand why there
are no HFQPOs seen when the accretion disk dominates the spectrum (red
x's in Figs.~\ref{fig:sp1550} and ~\ref{fig:sp1655}). Below we discuss
two types of inner disk oscillations, while noting that we cannot
exclude the possibility that high frequency oscillations are rooted in
the physics of the corona that is responsible for the power-law component.

\subsection{Resonance in GR Coordinate Frequencies in the Inner Disk}

At the radii in the accretion disk where most of the X-rays originate,
the coordinate frequencies in GR are predicted to have non-integral
ratios. The 3:2 frequency ratio in the HFQPOs in GRO~J1655--40 was
therefore seen by Abramowicz \& Kluzniak (2001) as remarkable support
for their idea that QPOs may result from a resonance between the
orbital (i.e., azimuthal) and radial coordinate frequencies in the
inner disk.  Unlike the orbital and polar frequencies, the radial
coordinate frequency reaches a maximum value at a radius larger than
the innermost stable orbit, regardless of the values of black hole
mass and spin (see \cite{kat01}; \cite{mer01}; and references
therein). For a wide range in the spin parameter, $a_*$, one can find
a particular radius that corresponds to a 1:2 or 1:3 ratio in the
radial and orbital coordinate frequencies. In the resonance model,
nonlinear perturbations may grow at these radii, ultimately producing
X-ray oscillations that represent some combination of the individual
resonance frequencies, their sum, or their difference. The proper
interpretation of observed HFQPO frequencies may then constrain the
black hole spin parameter, if the mass is known via dynamical studies
of the binary companion. For GRO~J1655--40, Abramowicz \& Kluzniak
(2001) constrain the spin parameter to the range $0.2 < a_* < 0.67$
for a black hole mass in the range 5.5 to 7.9 \msun.  For the same
limits in black hole mass and spin, the resonance radius occurs in the
range 4.1-7.2 $R_g$, where $R_g = G M / c^2$. We calculate that the
resonance radius in these cases lies in-between the innermost stable
orbit (3.5-5.3, $R_g$) and the radius of maximum surface emissivity
(5.4-8.4 $R_g$; \cite{mer01}; \cite{zha97}). Resonance may therefore
occur near radii already expected to yield X-ray emission.

For the case of XTE~J1550-564, a black hole mass of 8.5 to 11.5 \msun
(\cite{oro02}) can be combined with the HFQPO oscillations to
constrain the value of $a_*$ using the resonance model, as shown in
the top panel of Fig.~\ref{fig:res}. The blue tracks show values of
black hole mass and spin that produce orbital and radial coordinate
frequencies that are within 2\% of 184 and 92 Hz (left track),
respectively, or 276 and 92 Hz (right track), respectively. This model
yields limits on the black hole spin: $0.1 < a_* < 0.6$. At the lower
spin limit (8.5 \msun and the 2:1 resonance), the resonance radius is
7.5 $R_g$, while the innermost stable orbit is 5.6 $R_g$ and a maximum
surface emissivity (with no resonance) would occur at 8.8 $R_g$. At
the spin upper limit (11.5 \msun and a 3:1 resonance) the resonance
occurs at 4.5 $R_g$, with a last stable orbit of 3.9 $R_g$ and maximum
surface emissivity at 6.0 $R_g$.

In the lower panel of Fig.~\ref{fig:res}, we illustrate the resonance
tracks for GRO~J1655--40 for the range 5.8-6.8 \msun (\cite{gre01})
that correspond to orbital and radial coordinate frequencies within
2\% of 300 and 150 Hz (left track), respectively, and 450 and 150 Hz
(right track), respectively. The resonance model yields $0.24 < a_* <
0.58$ for GRO~J1655--40. The results of Abramowicz \& Kluzniak (2001)
differ slightly because they used broader mass constraints (5.5-7.9
\msun; \cite{sha99}).

In both X-ray sources the HFQPO detections occur with simultaneous
LFQPO detections, such as the 5.8 Hz QPO for the type A PDS and the
6.4 Hz QPO for type B PDS in XTE~J1550--564 shown in
Fig.~\ref{fig:pds1550}.  In the case of GRO~J1655--40, there is more
than one LFQPO present in some observations (\cite{rem99b}), and we
select the QPO at 12.2-17.5 Hz that is very strong in the highest
energy band, since it is present in all of the observations that
exhibit the HFQPO at either 300 or 450 Hz. We have examined all of the
coordinate frequencies and their beat frequencies at the resonance
radii consistent with the HFQPOs in each of the two sources. We
conclude that the 2:1 resonances, but not the 3:1 resonance, can
possibly account for the observed LFQPO frequencies as a beat between
the orbital and polar coordinate frequencies. This beat frequency is a
precession known as ``frame-dragging'' (e.g. \cite{mer01}). In the top
panel of Fig.~\ref{fig:res}, the regions shaded with darker blue show
the subset of resonance parameters for XTE~J1550--564 in which the
frame-dragging frequency lies in the range 5.8-6.4 Hz. Similarly for
GRO~J1655-40, in the bottom panel of Fig.~\ref{fig:res}, the darker
blue region shows black hole parameters for which the frame dragging
frequency is in the range 12.2-17.5 Hz.  We caution that the
identification of particular LFQPO features with relativistic frame
dragging is, at this point, highly speculative.

Thus far, there is no detailed description as to how a resonance
between the orbital and radial coordinate frequencies in GR would
produce the requisite X-ray oscillations and the alternating
conditions that cause the 2nd or 3rd harmonic to become the dominant
QPO. Perhaps even more challenging is the evidence for frequency
jitter in the HFQPOs of XTE~J1550-564. Because the radial dependence
of the two relevant coordinate frequencies is so different, the
resonances must be confined to a narrow range in radius. This is a
particular problem for the 3:1 resonance, where the ratio of the
orbital to radial coordinate frequencies changes rapidly with radius.
We illustrate the diverging coordinate frequencies
in Fig.~\ref{fig:res_width}.  in the case of a 10 \msun black hole
with $a_* = 0.465$ (top panel), a 276:92 Hz resonance occurs at 5.03
$r_g$; these mass and spin values are within the 3:1 resonance (right
track) shown in Fig.~\ref{fig:res} (top panel).  Here, a shift in the
orbital frequency by $\pm$10\% implies that the radius must vary
between 4.7 and 5.4 $r_g$, and the ratio of coordinate frequencies
would then vary in the range 2.5-4.1. It may then be impossible to
shift the radius of the perturbation enough to match the shifted
HFQPOs and yet maintain the perturbation (no longer in resonance)
throughout an RXTE observation, which is longer than the dynamical
time by a factor of $\sim 10^6$. The problem is less serious, but
nontrivial, for the 2:1 resonance (see Fig.~\ref{fig:res_width}): for
the same black hole mass and $a_* = 0.324$ (consistent with a 184:92
Hz resonance at 6.72 $r_g$), a 10\% frequency shift corresponds to a
range of radii of 6.3-7.2 $r_g$, over which the ratio of coordinate
frequencies changes by about $\pm 20$ \%.

Finally, we note that all of the QPO models
suffer from the ongoing uncertainty as to the origin and geometry of
the X-ray power-law component, widely attributed to inverse Compton
scattering of thermal photons.  We cannot posit a simple way in
which the resonance model can appear to shift harmonics in response to
increased Comptonization.

\subsection{Diskoseismic Oscillation Models}

A different model that warrants consideration with respect to harmonic
HFQPOs in black hole binaries is the diskoseismic model (\cite{kat80};
\cite{wag99}; \cite{kat01}; \cite{wag01}). Here, GR theory predicts
that the inner accretion disk may trap oscillations, which is a
consequence related to the turnover in the radial coordinate frequency
at small radii, which was noted previously.  The concept of a
resonance cavity in the inner disk is naturally attractive with
respect to our interpretation of harmonically related HFQPOs.  The
three dimensional character of the model allows for modest shifts in
oscillation frequency, e.g. with changes in disk thickness and
luminosity (\cite{wag99}).  Such changes in disk conditions would, in
principle, produce coupled changes in both the X-ray spectrum and the
oscillation frequency.

The strongest oscillations are expected to arise from gravity modes
(``g-modes''), which were investigated in Kerr geometry by Perez et
al. (1997). \nocite{per97} For adiabatic perturbations, the
eigenfunction solution predicts a fundamental radial mode ($m=0$) that
would be in the range $\sim 70$ to 110 Hz for a 10 \msun black hole
with $0 < a_* < 0.5$. However, for higher $m$-number, the g-mode
frequencies do not increase by integral values (\cite{per97}), and so
the predictions do not match the observed HFQPO frequencies. It was
noted that the ratio of the orbital frequency at the inner disk radius
to the fundamental g-mode frequency is close to a value of 3.08 for a
wide range in $a_*$ (\cite{wag99}).  However, this may suggest a
strong feature at $\nu$ (i.e. the g-mode oscillation) and a weak
feature near $3 \nu$, which does not resemble the observations.
Investigations have also been made for diskoseismic p-modes
(\cite{ort00}) and c-modes (\cite{sil01}), but neither study predicts
a system of linear harmonics. Kato (2001) has pointed out that the
results of numerical simulations and eigenmode analyses show some
differences, and therefore further study is warranted.

\acknowledgements Partial support for R.R. and M.M. was provided by
the NASA contract to M.I.T. for RXTE instruments. J.M. acknowledges
the support of NASA grant NAG5-10813. Radleigh Santos helped to
process RXTE data for the 2000 outburst of XTE~J1550-564. We thank
Michiel van der Klis and Al Levine for providing helpful comments.

\clearpage

\begin{deluxetable}{lccccc}
\tablenum{1}
\tablecaption{QPO Fit Parameters\tablenotemark{a}}
\small
\tablehead{
 \colhead{Parameter} &   \colhead{units}   &    \colhead{1998-99 A}    &    \colhead{1998-99 B}   &   \colhead{1998 Sep 19}  &   \colhead{2000 Apr 30-May 9}
}
\startdata
No. of Obs. & \nodata &  10  &    9   &    1   &    12  \\
Source Intensity  &  Crab units  &  0.3--2.0    &  1.3--1.8    &  6.5    &  0.3--1.0 \\
$\chi_{\nu}^2$ & \nodata &    0.63   &   0.72  &  1.10 &  0.92 \\ \\

Frequency  &    Hz    & 281.7 (1.5) & 278.0  & \nodata & 270.3 (3.0) \\
Significance & $\sigma$ &    16.8     &  4.0    &  \nodata  &  15.2 \\
FWHM      &    Hz    & 33.6 (2.8)  &  38.6 (12.0) & \nodata & 50.0 (8.0) \\
Q ($\nu/FWHM$) & \nodata & 8.4 (0.5)  & 7.2 (1.3) & \nodata & 5.4 (1.0) \\
Amplitude & rms \% & 3.12 (0.09) & 0.70 (0.09) & \nodata & 5.05 (0.17) \\ \\

Frequency  &    Hz    & 187.8  & 185.3 (3.7) & 189.6 (5.1) & 180.22  \\
Significance & $\sigma$ &     3.9     &    13.7      &  6.6  & 5.0 \\
FWHM      &    Hz    & 22.0 (6.0)  &  66.1 (10.3) & 49.4 ( 10.4) & 32.0 (5.0) \\
Q ($\nu/FWHM$) & \nodata & 8.5 (1.6) & 2.8 (0.3) & 3.8 (0.6) & 5.6 (0.8) \\
Amplitude & rms \% & 1.19 (0.15) & 1.07 (0.08) & 1.18 (0.09) & 2.27 (0.23) \\ \\

Frequency  &    Hz   & \nodata & 92.7  & \nodata & \nodata \\
Significance & $\sigma$ & \nodata   &  4.1     & \nodata & \nodata \\
FWHM      &    Hz    & \nodata & 30.4 (6.0) & \nodata & \nodata \\
Q ($\nu/FWHM$) & \nodata & \nodata  & 3.0 (0.4) & \nodata & \nodata \\
Amplitude & rms \% & \nodata & 0.67 (0.08) & \nodata & \nodata \\ \\

\enddata

\tablenotetext{a} {Uncertainty estimates (1 $\sigma$ confidence) are
given in parentheses.  Since the model includes only one free
parameter for frequency, the uncertainty in the frequency is listed
for the harmonic that dominates the fit.  All of the QPO results
pertain to the spectral range 6 to 30 keV, while the source intensity
is evaluated at 2-10 keV, where 1 Crab $= 2.4 \times 10^{-8}$ erg
cm$^{-2}$ s$^{-1}$. }

\end{deluxetable}

\figcaption[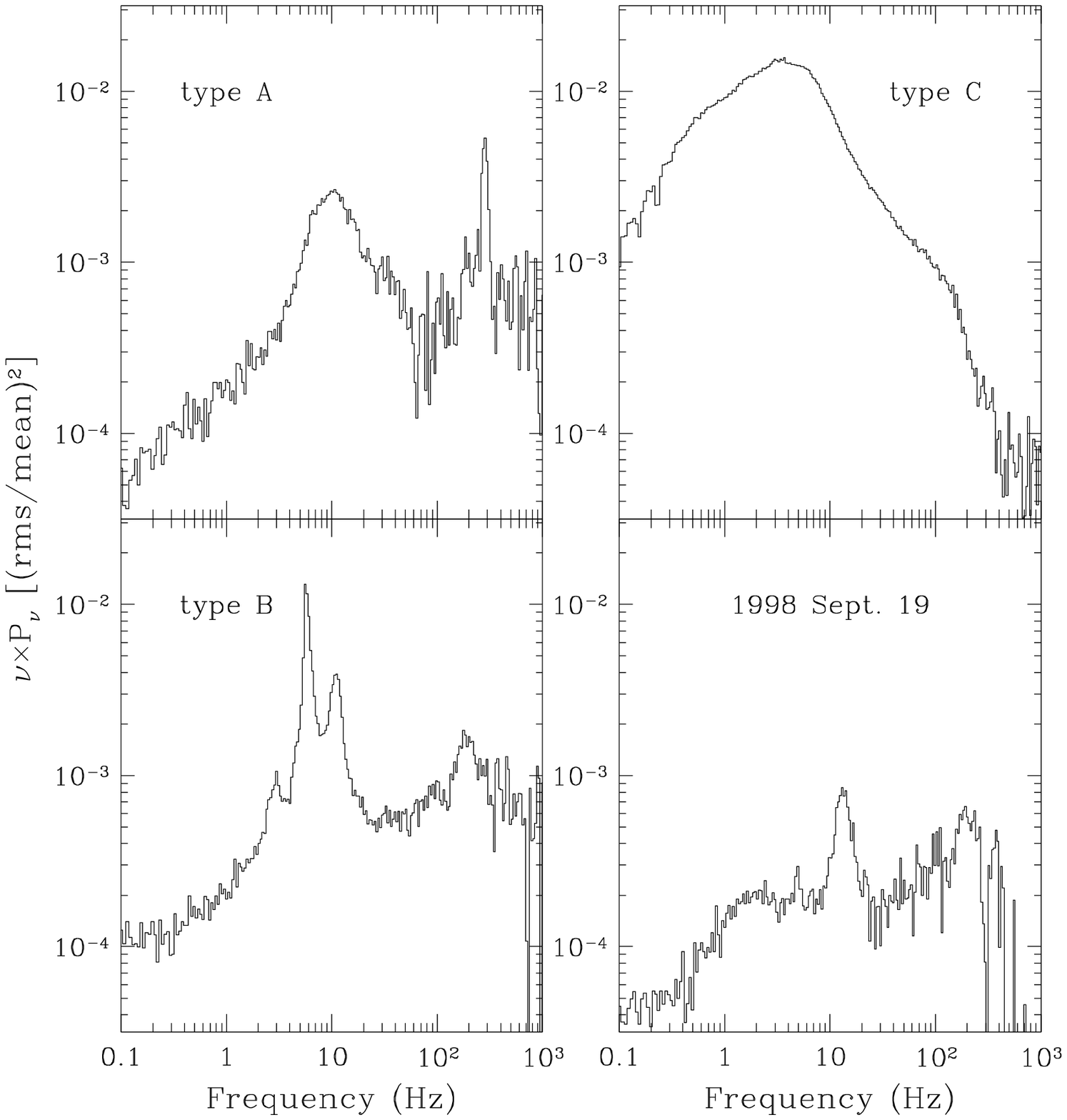]{Average power spectra for observations of
XTE~J1550--564 grouped by the type (A, B, or C) of low-frequency QPOs
that were observed during the 1998-1999 outburst. We also show the PDS
for the intense 7 Crab flare that was sampled on 1998 September
19. \label{fig:pds1550}}

\figcaption[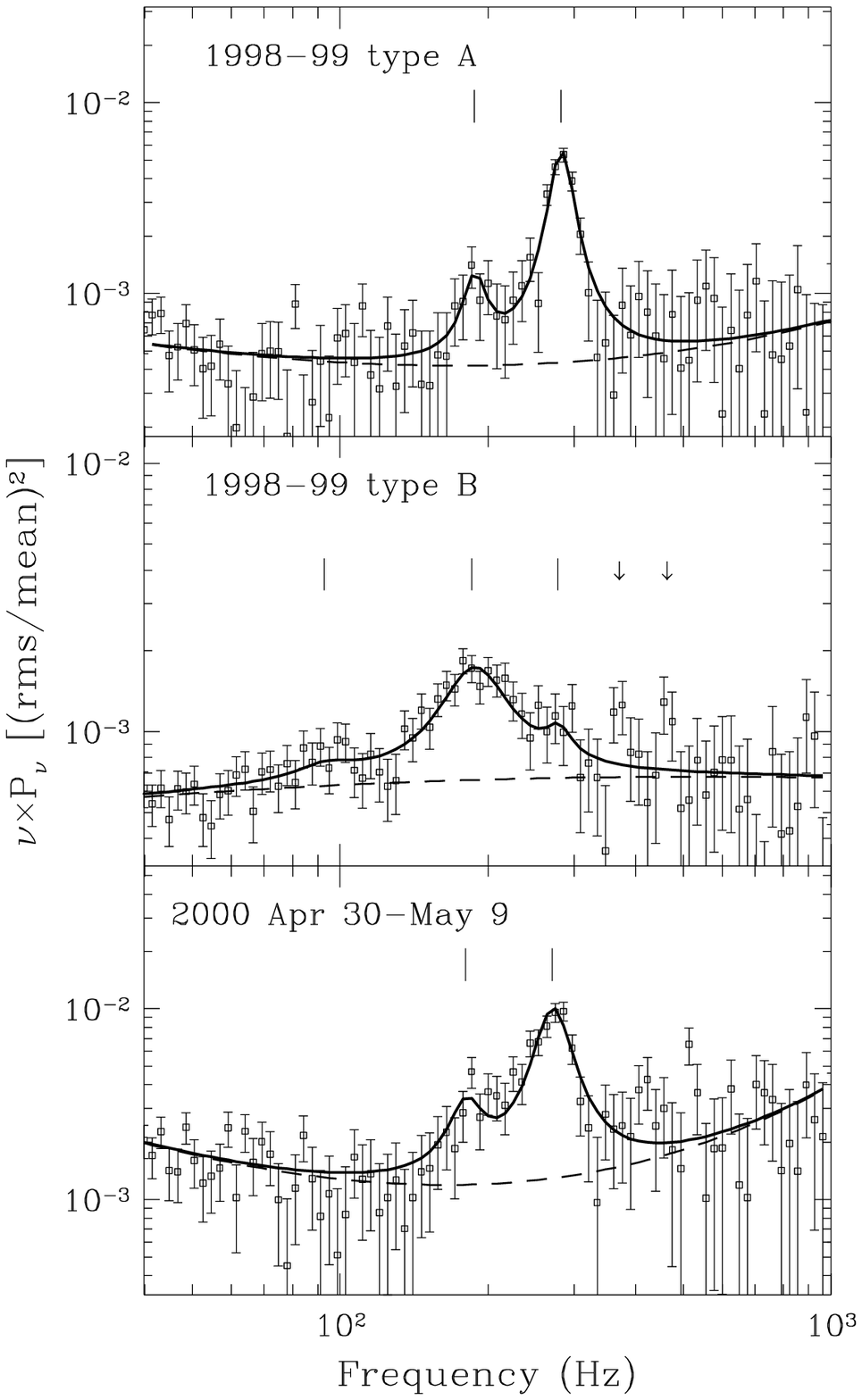]{Fits for harmonically related HFQPOs in
XTE~J1550--564 in power spectra at 6--30 keV.  The top two
panels show the same data displayed in Fig. 1 (1998-1999 averages for
LFQPO types A and B). The bottom panel shows the QPO fit for the
average of 12 observations between 2000 April 30 and May 9. In
each panel, the tick marks above the data show the central frequencies
of significant QPOs. The best fit is shown with a smooth, dark
curve, and the power continuum is shown with a dashed line. For the
type B group (middle panel), the arrows show the expected locations of
the 4th and 5th harmonics. \label{fig:mdl1550}}

\figcaption[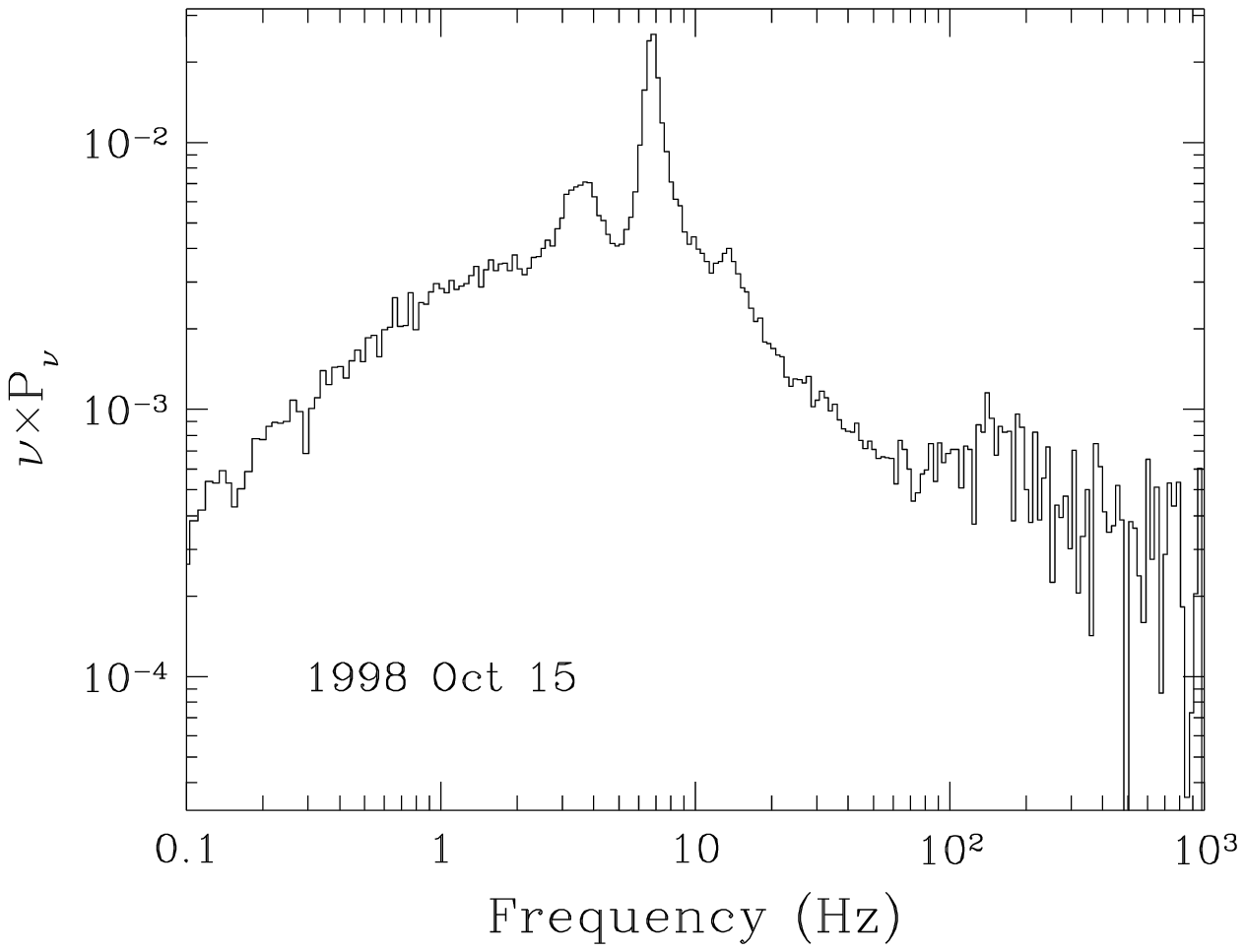]{Histogram of HFQPO frequencies for
XTE~J1550--564 and GRO~J1655--40. The binning intervals 
are varied to maintain a width of $\pm$5 \% relative to the
central frequency. Each source displays two peaks in the distribution
that have a 3:2 ratio in frequency. \label{fig:hist}}

\figcaption[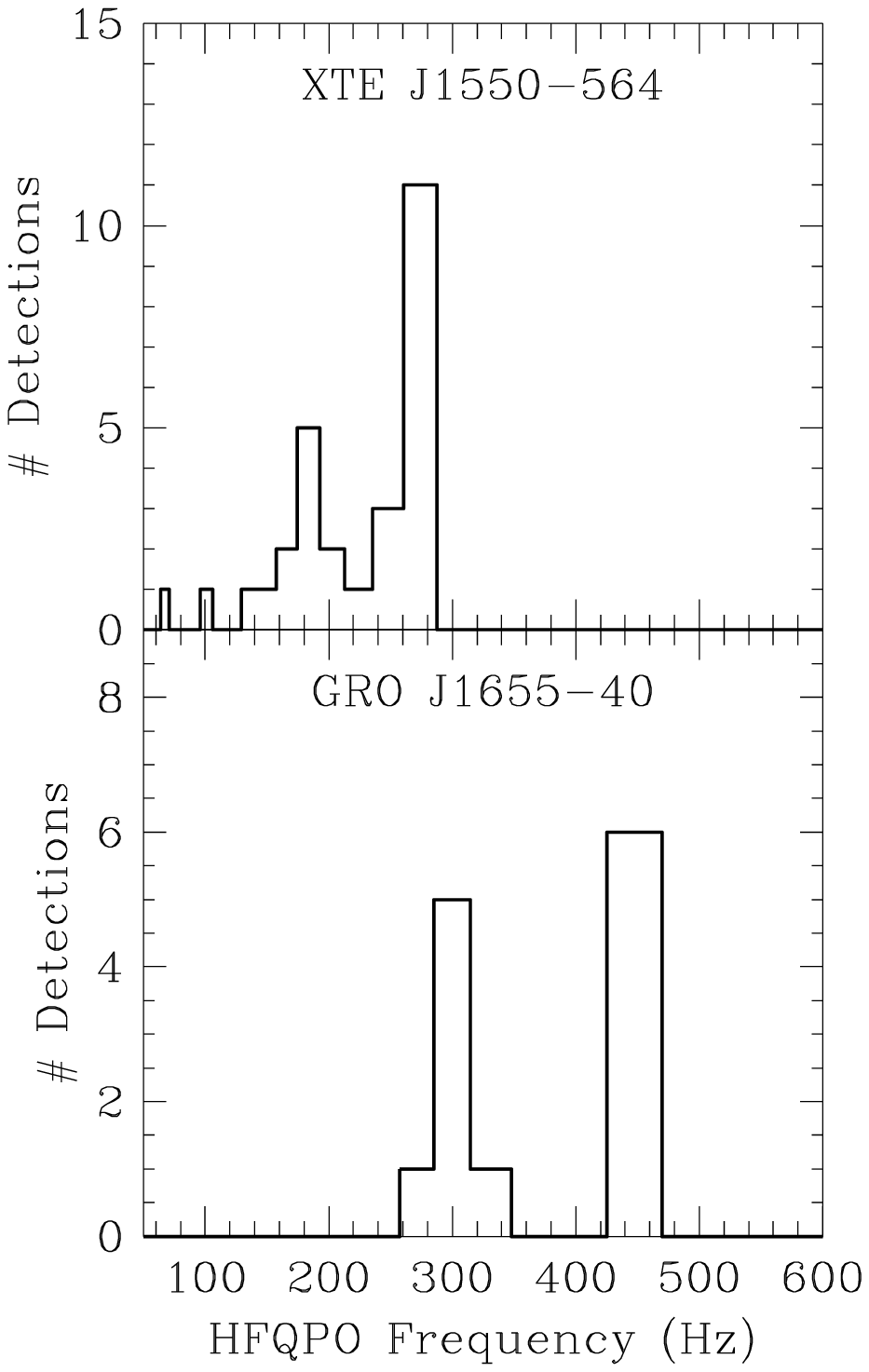]{Average power spectrum for XTE~J1550--564 for
the two observations made on 1998 October 15. The QPO at $143 \pm 8$
Hz represents the most significant deviation from the harmonic scheme
(92, 184, 276 Hz) that accounts for most of the HFQPO detections in
this source. \label{fig:oct15}}

\figcaption[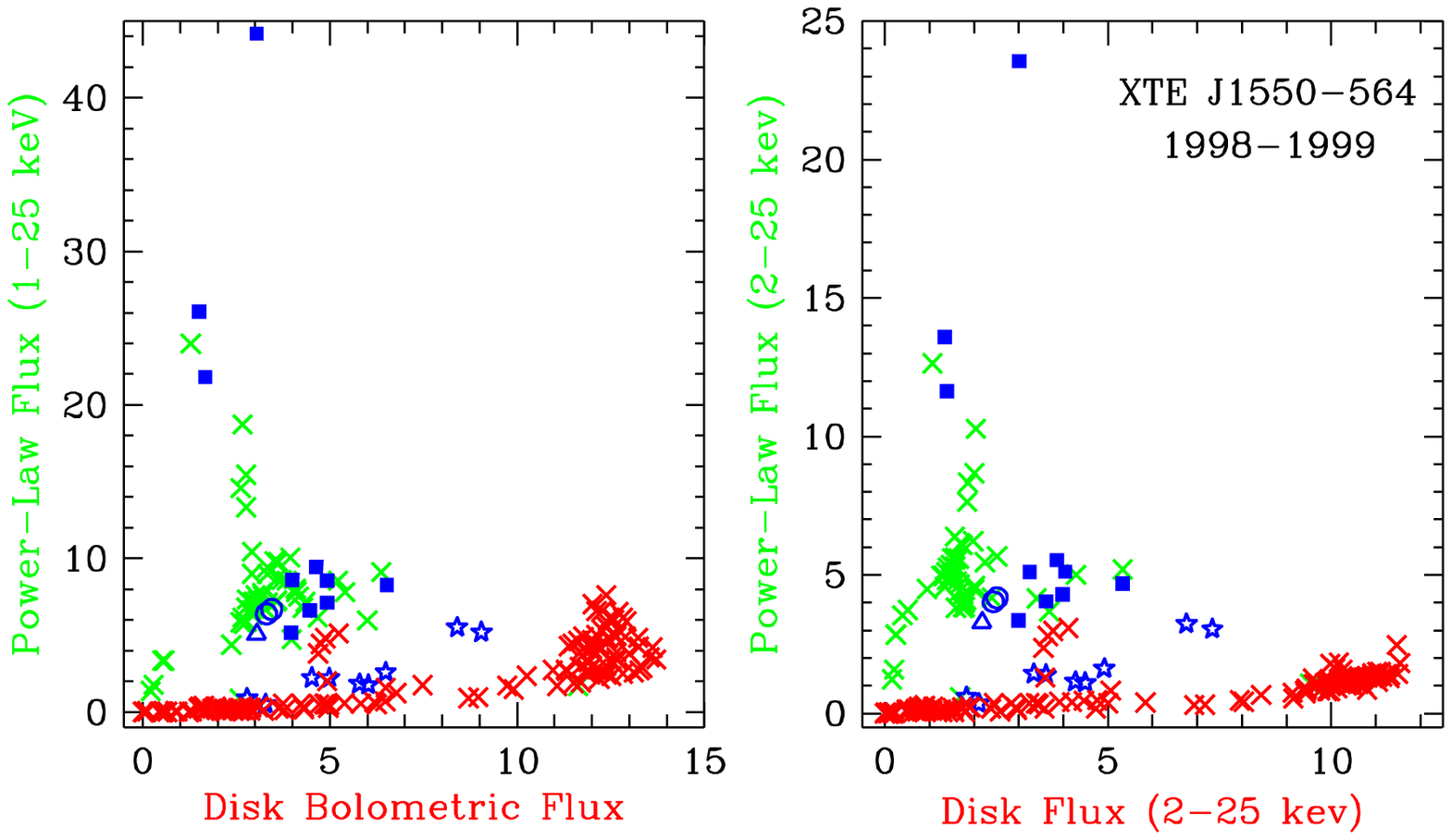]{Energy spectrum decomposition for XTE~J1550--564
during the 1998-1999 outburst. The flux units are $10^{-8}$ erg
cm$^{-2}$ s$^{-1}$. The flux from the accretion disk and the X-ray
power law are shown in conventions of both bolometric flux
(extrapolated; left) and apparent flux (2-25 keV; right). The color of
the plotting symbol denotes the QPO conditions: HFQPO detections
(blue), only LFQPOs (green ``x''), and no QPOs (red ``x'').  In
addition, the shape of the blue symbols distinguishes the HFQPOs near
92 Hz (triangle), 184 Hz (filled square), and 276 Hz (star). The pair
of observations on 1998 October 15 are displayed with blue circles. In
either flux convention (bolometric or apparent), there is a systematic
shift away from disk-dominated track (red x's) as the detected HFQPO
shifts from 276 to 184 Hz. \label{fig:sp1550}}

\figcaption[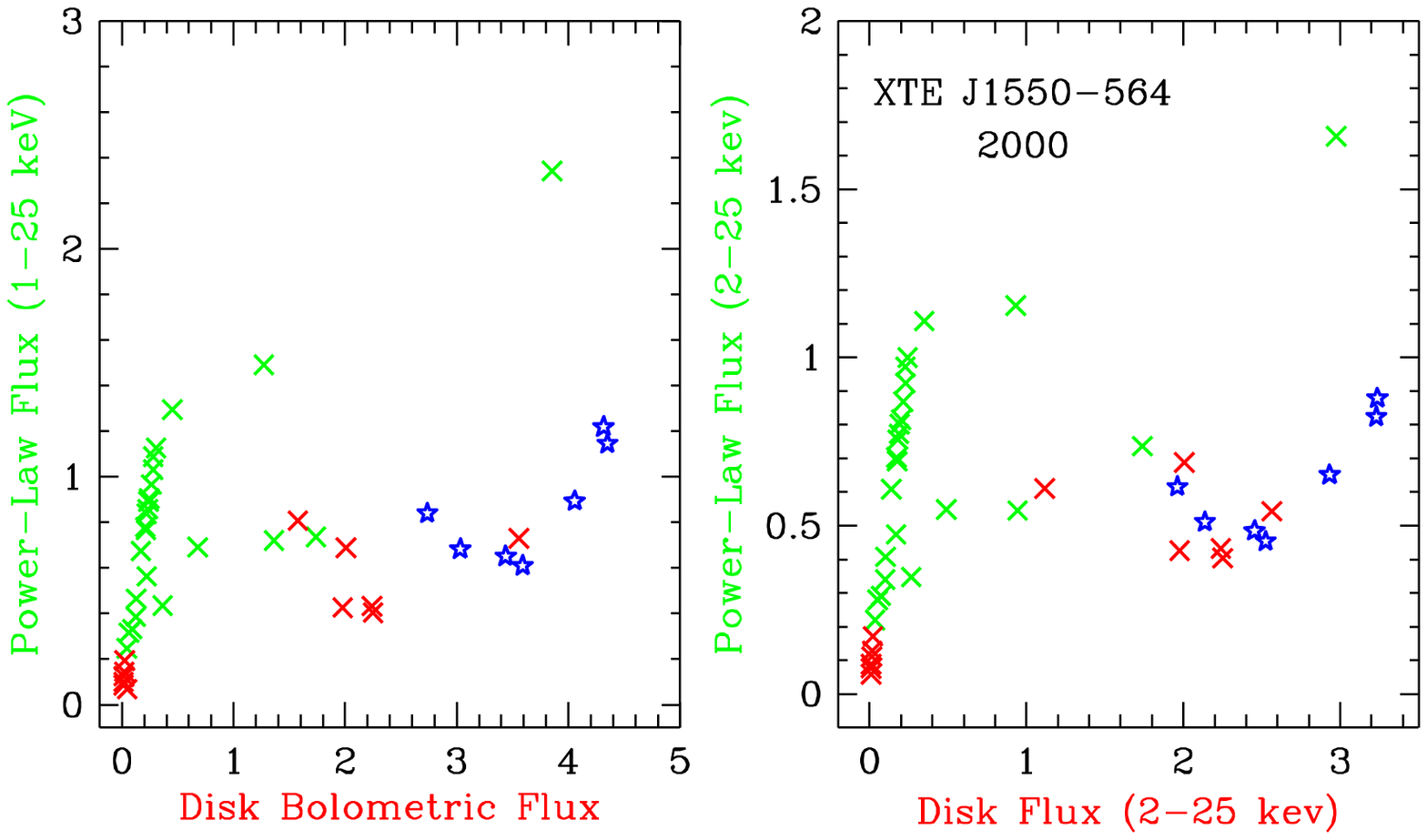]{Energy spectrum decomposition for
XTE~J1550--564 during the outburst of 2000. The flux units are
$10^{-8}$ erg cm$^{-2}$ s$^{-1}$. The symbols follow the
definitions given for Fig. 4. In this outburst there is a track in
which the power-law spectrum dominates (green x's), and the points
associated with HFQPOs near 276 Hz (blue stars) are again close to the
disk-dominated track shown in Fig. 4. \label{fig:sp1550b}}

\figcaption[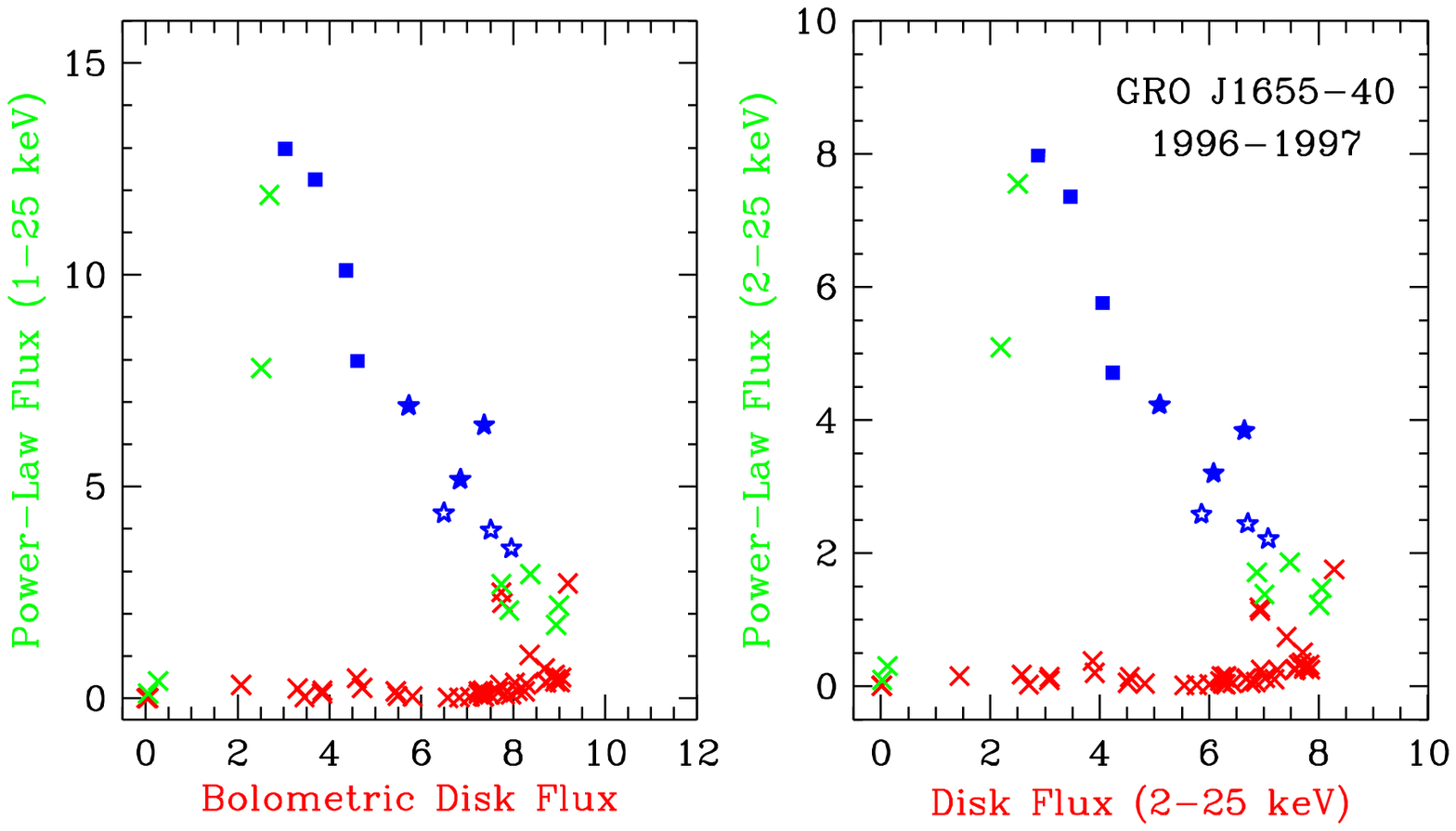]{Energy spectrum decomposition for
GRO~J1655--40 during its outburst of 1996-1997. The flux units are
$10^{-8}$ erg cm$^{-2}$ s$^{-1}$. The symbols follow the
definitions given for Fig. 4, except that the observations that
exhibit both HFQPOs (300 and 450 Hz) are shown with solid stars. The
results are very similar to those for XTE~J1550--564 (Fig. 4). There is
increasing power-law flux as the strongest HFQPO shifts from the 450
Hz feature (open stars) to the 300 Hz feature (solid
squares). \label{fig:sp1655}}

\figcaption[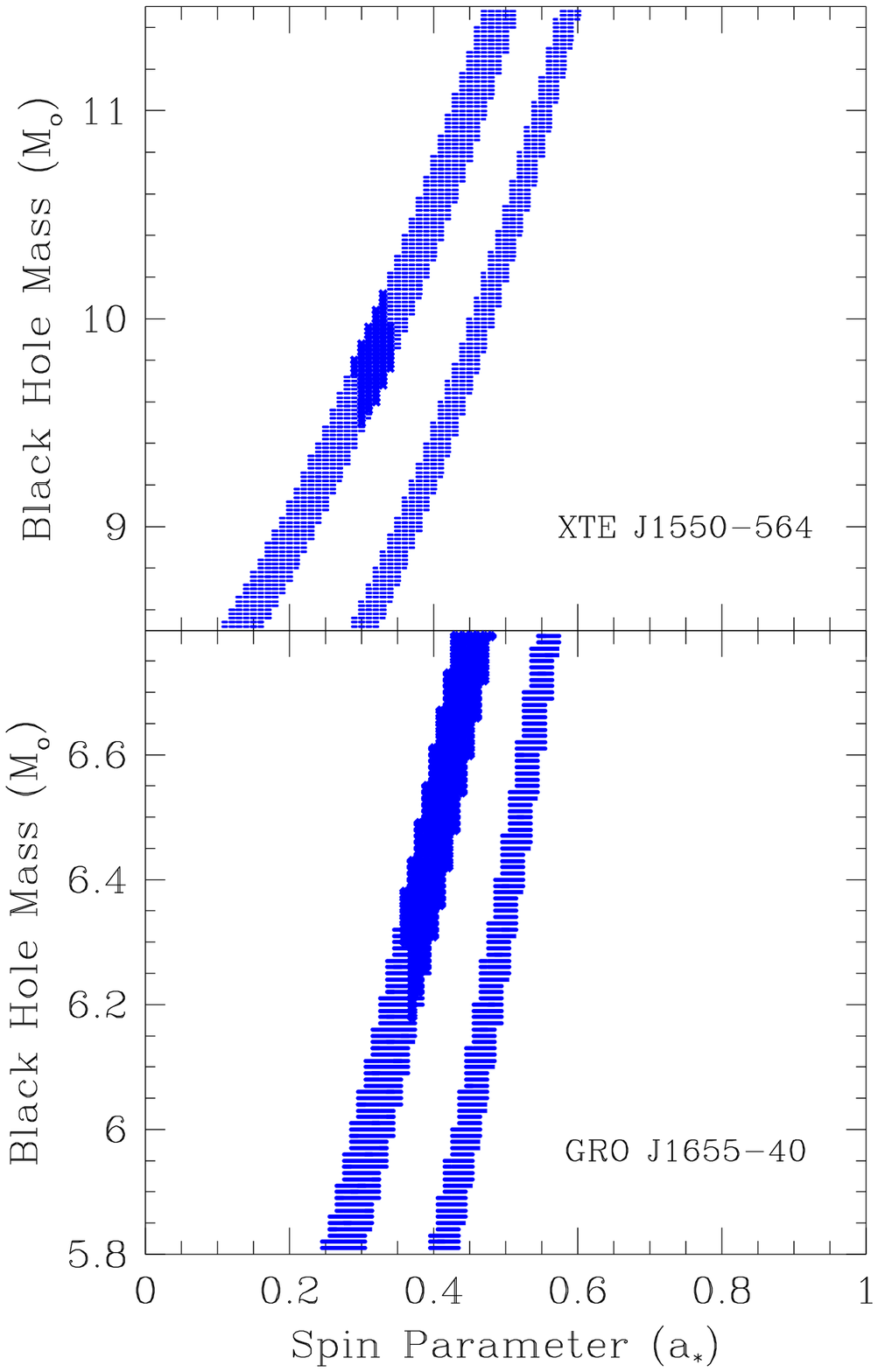]{Application of the coordinate-frequency resonance
model to the HFQPOs in XTE~J1550--564 (top) and GRO~J1655--40
(bottom). At each point in the blue shaded regions, the orbital and
radial frequencies are within 2\% of the HFQPO harmonic
frequencies. In each panel, the left track corresponds with the 2:1
resonance (i.e. 184 : 92 Hz in the top panel and 300 : 150 Hz in the
bottom panel), and the right track corresponds with the 3:1 resonance
(276 : 92 Hz and 450 : 150 Hz, respectively).  The regions shaded with
darker blue show the parameters that additionally yield a
frame-dragging frequency in the range of 5.8-6.4 Hz (top) and
12.2-17.5 Hz (bottom).  We caution that the association of LFQPOs with
the frame-dragging frequency is highly speculative. \label{fig:res}}

\figcaption[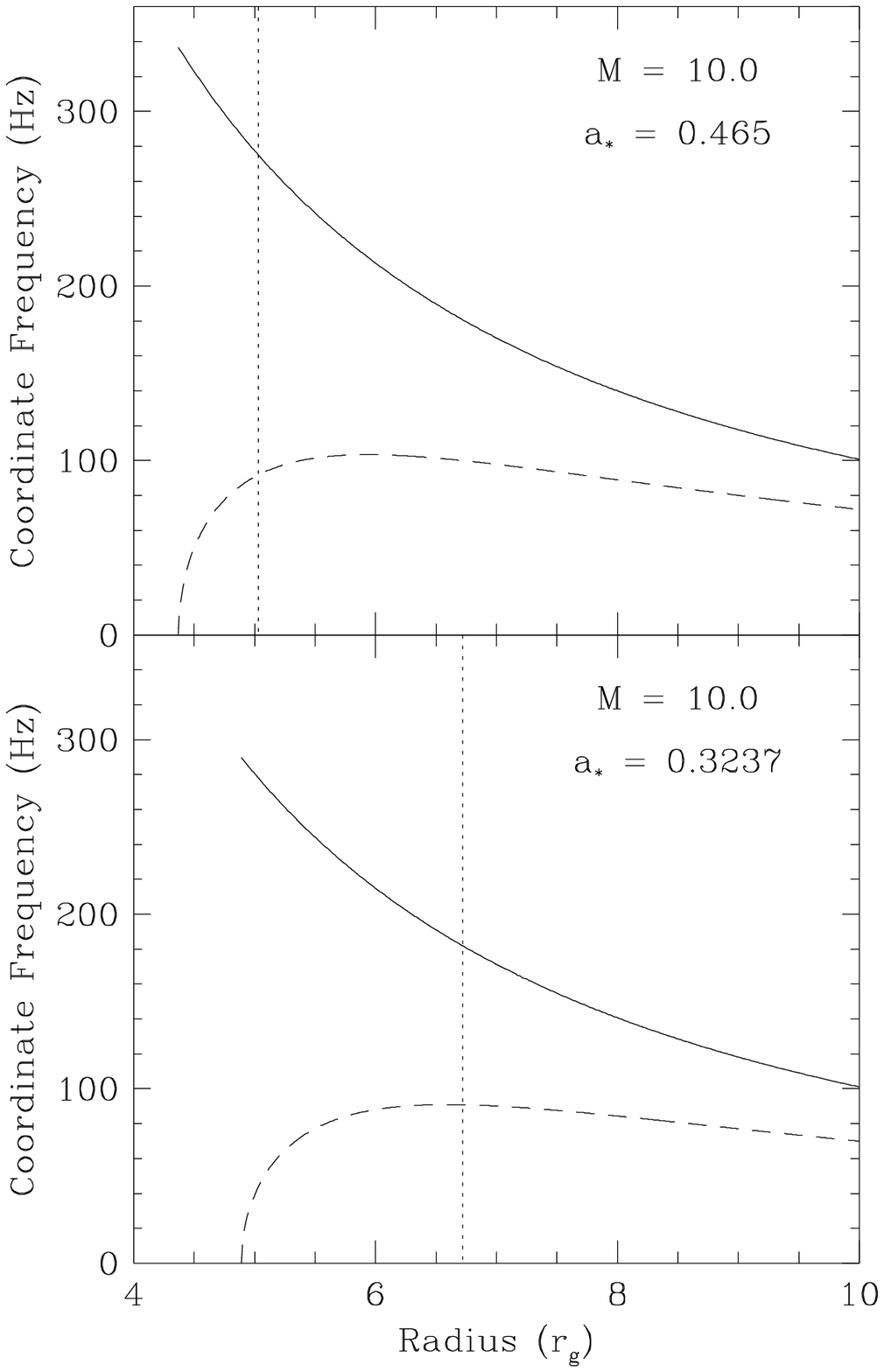]{Frequencies predicted by GR as a
function of radius, for cases that illustrate a 3:1 resonance (top)
and a 2:1 resonance (bottom) between the orbital (solid curve) and
radial (dashed curve) coordinate frequencies. In each case the mass is
10 \msun and the spin parameter has been chosen to match the
resonances to the values 276:92 Hz and 184:92 Hz, respectively. In
each case, the gradients in the coordinate frequencies have opposite
signs, and this creates a narrow region in radius where the frequency
ratio is near an integral value. Shifts in the observed frequencies
may therefore be very difficult to accomodate in the resonance model,
especially for the 3:1 resonance. \label{fig:res_width}}

\newpage
\begin{figure}
\figurenum{1} \plotone{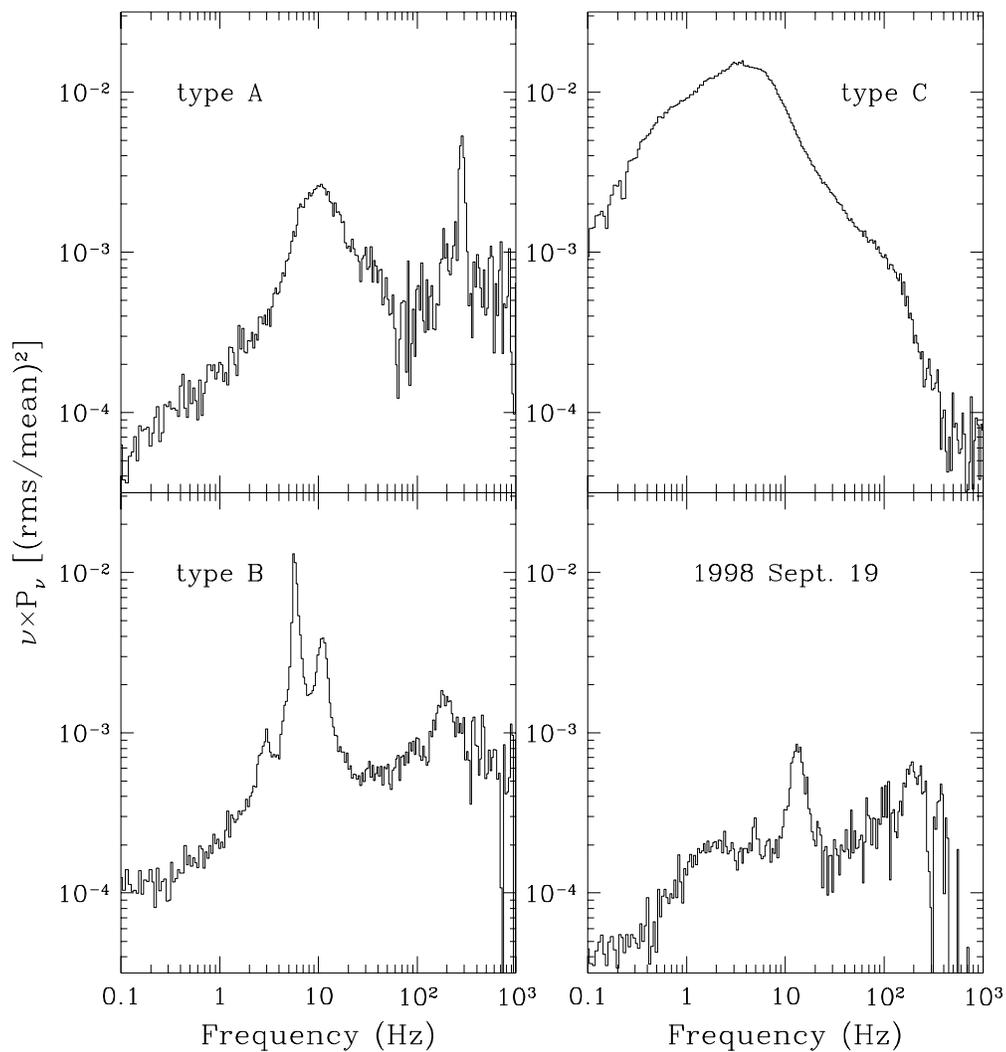}
\caption{Average power spectra for observations of XTE~J1550--564
grouped by the type (A, B, or C) of low-frequency QPOs that were
observed during the 1998-1999 outburst. We also show the PDS for the
intense 7 Crab flare that was sampled on 1998 September 19.}
\end{figure}

\newpage
\begin{figure}
\figurenum{2}
\plotone{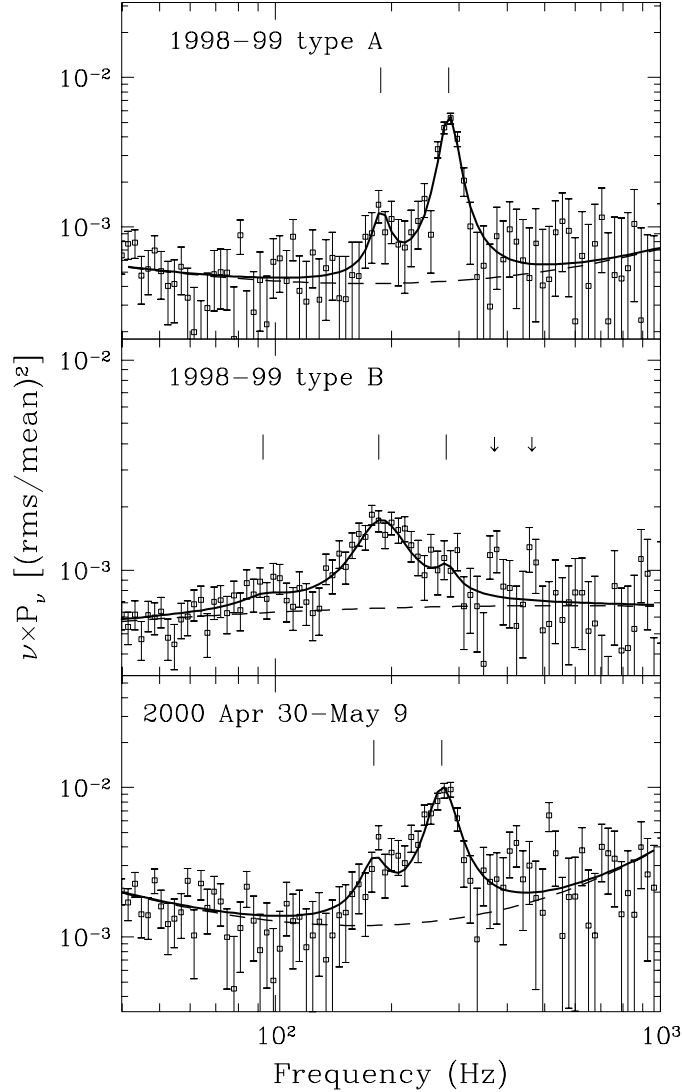}
\caption{Fits for harmonically related HFQPOs in
XTE~J1550--564 applied to power spectra at 6--30 keV.  The top two
panels show the same data displayed in Fig. 1 (1998-1999 averages for
LFQPO types A and B). The bottom panel shows the QPO fit for the
average of 12 observations made between 2000 April 30 and May 9. In
each panel, the tick marks above the data show the central frequencies
of the significant QPOs. The best fit is shown with a smooth, dark
curve, and the power continuum is shown with a dashed line. For the
type B group (middle panel), the arrows show the expected locations of
the 4th and 5th harmonics.}
\end{figure}

\newpage
\begin{figure}
\figurenum{3}
\plotone{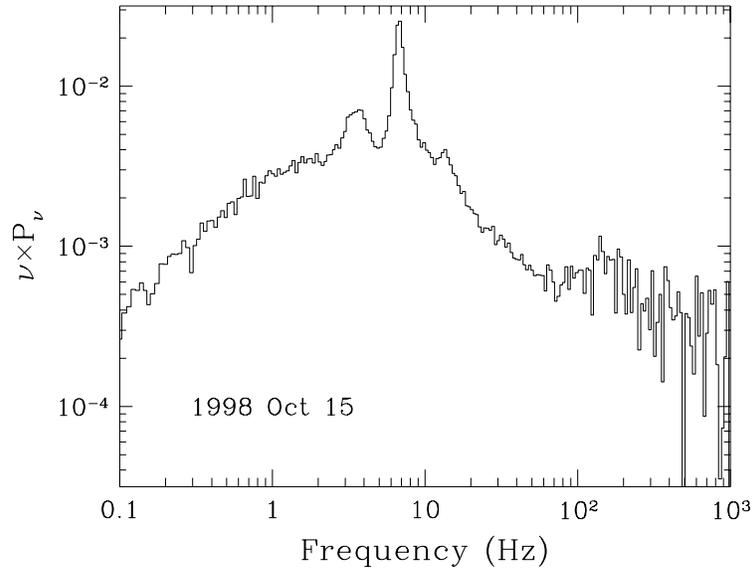}
\caption{Histogram of HFQPO frequencies for
XTE~J1550--564 and GRO~J1655--40. The binning intervals 
are varied to maintain a width of $\pm$5 \% relative to the
central frequency. Each source displays two peaks in the distribution
that have a 3:2 ratio in frequency.}
\end{figure}

\newpage
\begin{figure}
\figurenum{4}
\plotone{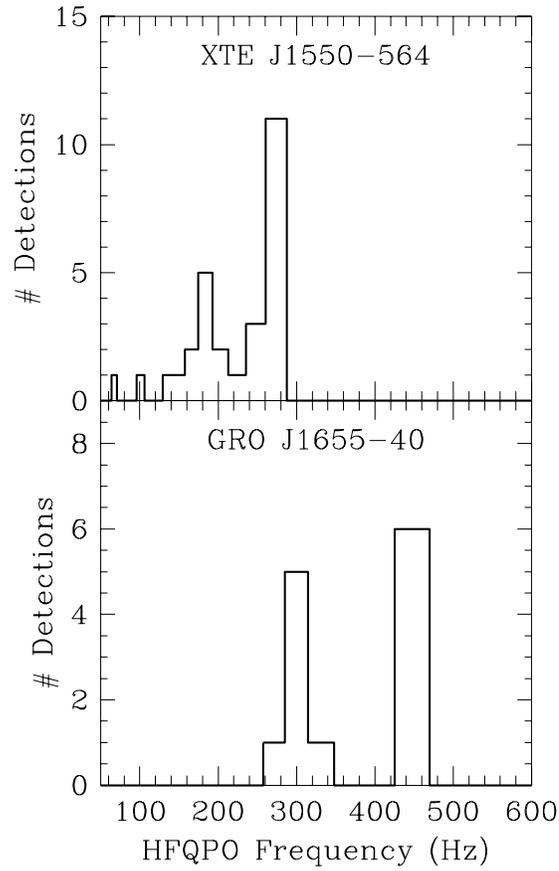}
\caption{Average power spectrum for XTE~J1550--564 for
the two observations made on 1998 October 15. The QPO at $143 \pm 8$
Hz represents the most significant deviation from the harmonic scheme
(92, 184, 276 Hz) that accounts for most of the HFQPO detections in
this source.}
\end{figure}

\newpage
\begin{figure}
\figurenum{5}
\plotone{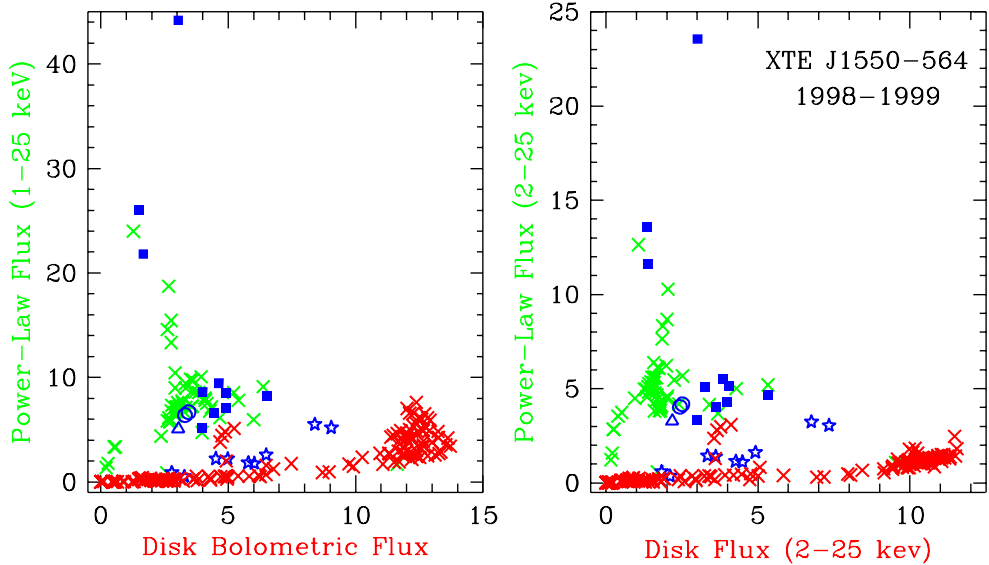}
\caption{Energy spectrum deconvolution for
XTE~J1550--564 during the 1998-1999 outburst.  The flux from the
accretion disk and the X-ray power law are shown in conventions of
both bolometric flux (extrapolated; left) and apparent flux (2-25 keV;
right). The color of the plotting symbol denotes the QPO conditions:
HFQPO detections (blue), only LFQPOs (green ``x''), and no QPOs (red
``x'').  In addition, the shape of the blue symbols distinguishes the
HFQPOs near 92 Hz (triangle), 184 Hz (filled square), and 276 Hz
(star). The pair of observations on 1998 October 15 are displayed with
blue circles. In either flux convention, there is a systematic shift
away from disk-dominated track (red x's) as the detected HFQPO shifts 
from 276 to 184 Hz.}
\end{figure}

\newpage
\begin{figure}
\figurenum{6}
\plotone{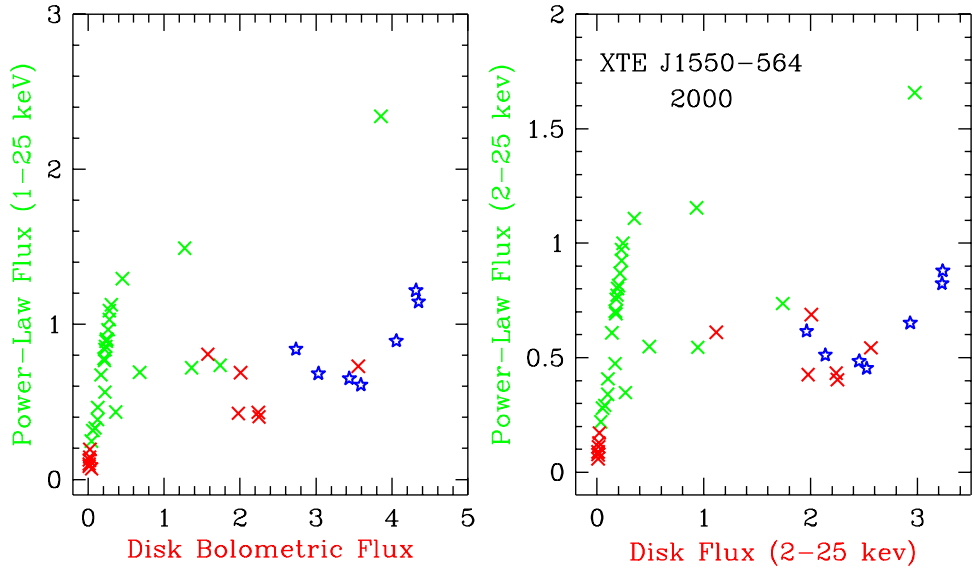}
\caption{Energy spectrum deconvolution for
XTE~J1550--564 during the outburst of 2000. The symbols follow the
definitions given for Fig. 4. In this outburst there is a track in
which the power-law spectrum dominates (green x's), and the points
associated with HFQPOs near 276 Hz (blue stars) are again close to the
disk-dominated track shown in Fig. 4.}
\end{figure}

\newpage
\begin{figure}
\figurenum{7}
\plotone{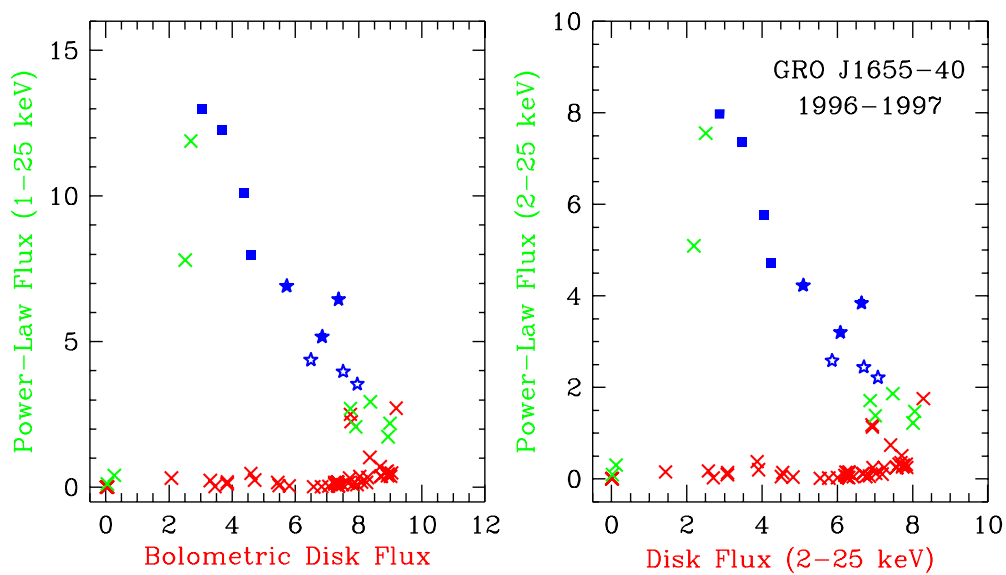}
\caption{Energy spectrum deconvolution for
GRO~J1655--40 during its outburst of 1996-1997. The symbols follow the
definitions given for Fig. 4, except that the observations that
exhibit both HFQPOs (300 and 450 Hz) are shown with solid stars. The
results are very similar to those for XTE~J1550--564 (Fig. 4). There is
increasing power-law flux as the strongest HFQPO shifts from the 450
Hz feature (open stars) to the 300 Hz feature (solid
squares).}
\end{figure}

\newpage
\begin{figure}
\figurenum{8}
\plotone{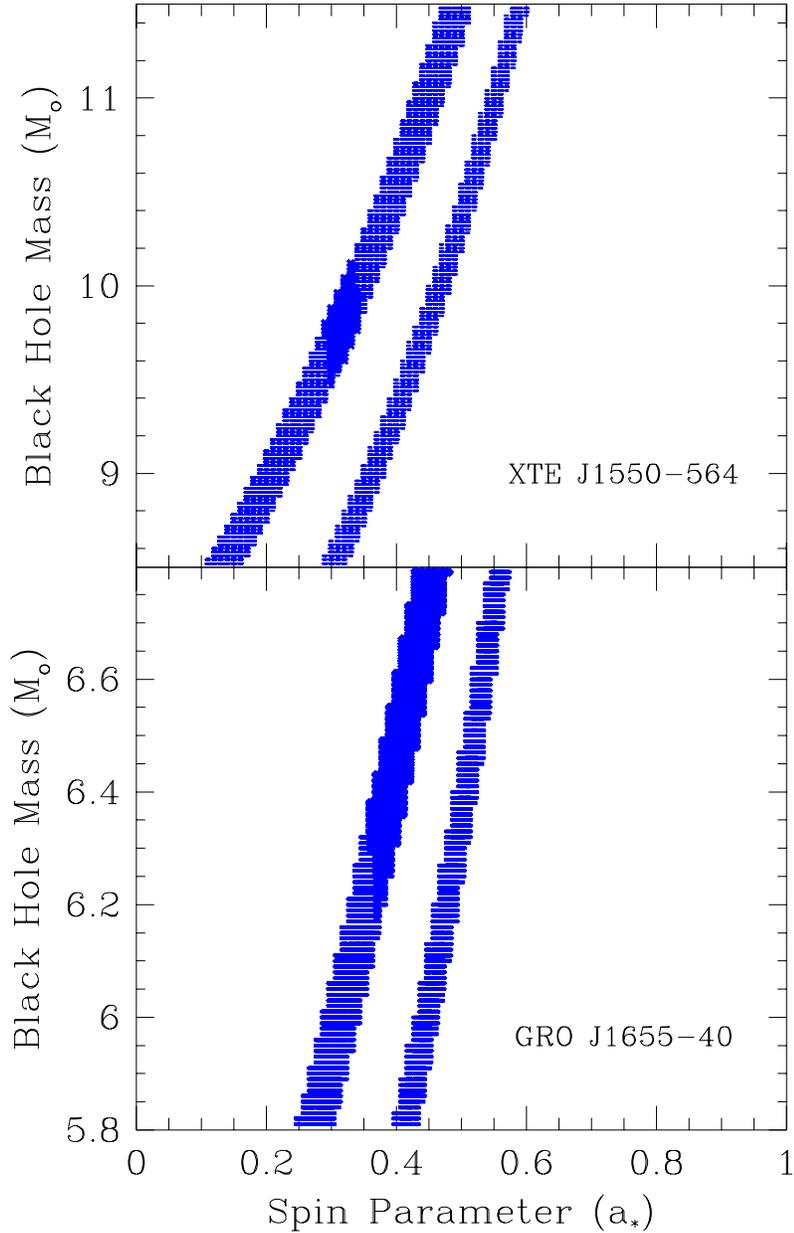}
\caption{Application of the coordinate-frequency resonance model to
the HFQPOs in XTE~J1550--564 (top) and GRO~J1655--40 (bottom). At each
point in the blue shaded regions, the orbital and radial frequencies
are within 2\% of the HFQPO harmonic frequencies. In each panel, the
left track corresponds with the 2:1 resonance (i.e. 184 : 92 Hz in the
top panel and 300 : 150 Hz in the bottom panel), and the right track
corresponds with the 3:1 resonance (276 : 92 Hz and 450 : 150 Hz,
respectively).  The regions shaded with darker blue show the
parameters that additionally yield a frame-dragging frequency
in the range of 5.8-6.4 Hz (top) and 12.2-17.5 Hz (bottom).
 We caution that the association of LFQPOs with the 
frame-dragging frequency is highly speculative.}
\end{figure}

\newpage
\begin{figure}
\figurenum{9}
\plotone{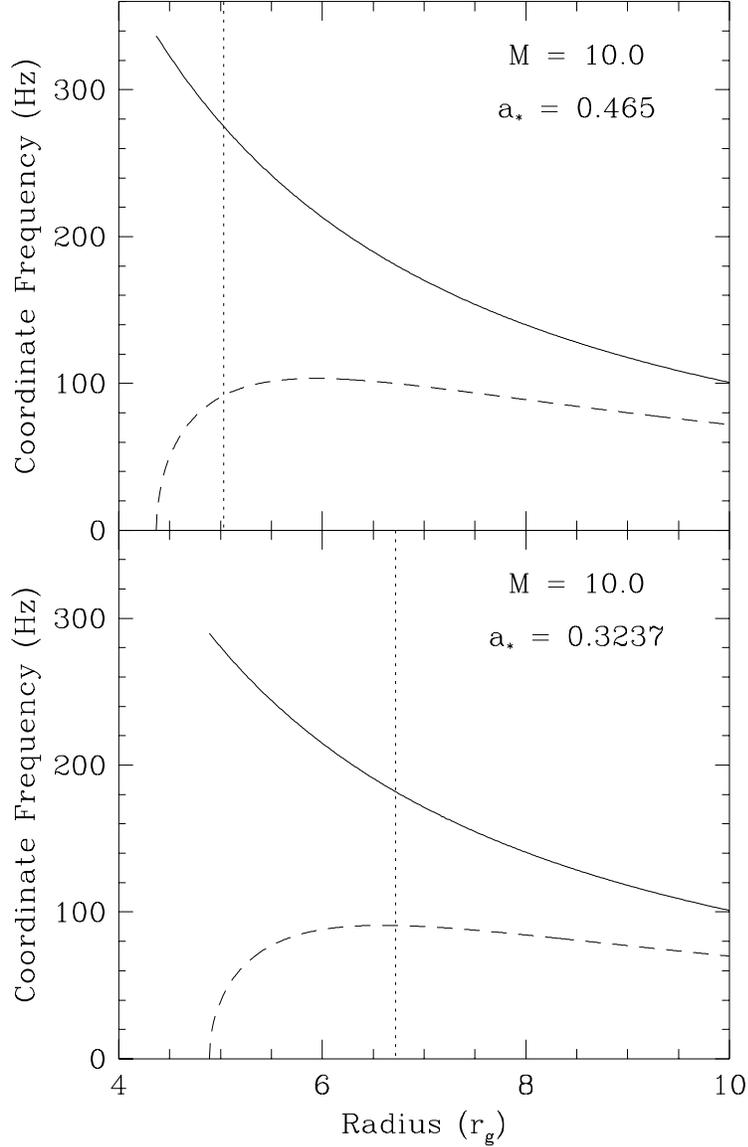}
\caption{Frequencies predicted by GR as a
function of radius, for cases that illustrate a 3:1 resonance (top)
and a 2:1 resonance (bottom) between the orbital (solid curve) and
radial (dashed curve) coordinate frequencies. In each case the mass is
10 \msun and the spin parameter has been chosen to match the
resonances to the values 276:92 Hz and 184:92 Hz, respectively. In
each case, the gradients in the coordinate frequencies have opposite
signs, and this creates a narrow region in radius where the frequency
ratio is near an integral value. Shifts in the observed frequencies
may therefore be very difficult to accomodate in the resonance model,
especially for the 3:1 resonance.}
\end{figure}


\begin{thebibliography}{}

\bibitem[Abramowicz \& Kluzniak 2001]{abr01}Abramowicz, M. A., \& Kluzniak, W., 2001, A\&A, 374, L19

\bibitem[Corbel et al. 2001]{cor01}Corbel, S. et al. 2001, \apj, 554, 43

\bibitem[Greene, Bailyn, \& Orosz 2001]{gre01}Greene, J., Bailyn, C. D., \& Orosz, J. A. 2001, \apj, 554, 1290

\bibitem[Hannikainen et al. 2001]{han01}Hannikainen, D., Wu, K.,
Campbell-Wilson, D., Hunstead, R., McIntyre, V., Lovell, J., Reynolds,
J., Tzioumis, T., \& Wu, K, 2001, Ap\&SS, 276, 45

\bibitem[Homan et al. 2001]{hom01}Homan, J., Wijnands, R., van der
Klis, M., Belloni, T., van Paradijs, J., Klein-Wolt, M., Fender, R.,
\& M\'{e}ndez, M. 2001, \apjs, 132, 377

\bibitem[Jain et al. 2001]{jai01}Jain, R., Bailyn, C. D., Orosz, J. A.,
McClintock, J. E., Sobczak, G. J., \& Remillard, R. A. 2001, \apj, 546, 1086

\bibitem[Kalemci et al. 2001]{kal01}Kalemci, E., Tomsick, J. A., Rothschild, R. E., Pottschmidt, K., \& Kaaret, P. 2001, \apj, 563, 239

\bibitem[Kato 2001]{kat01}Kato, S. 2001, PASJ, 53, 1

\bibitem[Kato \& Fukue 1980]{kat80}Kato, S., \& Fukue, J. 1980, PASJ, 32, 377

\bibitem[Leahy et al. 1983]{lea83}Leahy, D., Darbro, W., Weisskopf, M. C., Kahn, S., Sutherland, P. G., \& Grindlay, J. E. 1983, \apj, 266, 160

\bibitem[Merloni et al. 2001]{mer01}Merloni, A., Vietri, M., Stella, L., \& bini, D. 2001, \mnras, 304, 155

\bibitem[Miller et al. 2001]{mil01}Miller, J. M., et al. 2001, \apj, 563, 928

\bibitem[Morgan, Remillard, \& Greiner 1997]{mor97}Morgan, E. H.,
Remillard, R. A., \& Greiner, J. 1997, \apj, 482, 993

\bibitem[Muno, Morgan, \& Remillard 1999]{mun99} Muno, M. P., Morgan, E. H., \& Remillard, R. A. 1999, \apj, 527, 321

\bibitem[Nowak 2000]{now00}Nowak, M. 2000, \mnras, 318, 361

\bibitem[Orosz et al. 2002]{oro02}Orosz, J. A. et al. 2002; \apj, in press; astro-ph/0112101

\bibitem[Ortega-Rodriguez \& Wagoner 2000]{ort00}Ortega-Rodriguez, M., \& Wagoner, R. V. 2000, \apj, 543, 1060

\bibitem[Perez et al. 1997]{per97}Perez, C.A., Silbergleit, A.S., Wagoner, R.V., \& Lehr, D.E. 1997, \apj, 476 589 

\bibitem[Pottschmidt et al. 2002]{pot02}Pottschmidt, K., et al. 2002, submitted to A\&A

\bibitem[Psaltis, Belloni, \& van der Klis 1999]{psa99}Psaltis, D., Belloni, T., \& van der Klis, M. 1999, \apj, 520, 262 

\bibitem[Remillard et al. 1999a]{rem99a}Remillard, R. A., McClintock,
J. E., Sobczak, G. J., Bailyn, C. D., Orosz, J. A., Morgan, E. H., \&
Levine, A. M. 1999a, \apj, 517, L127

\bibitem[Remillard et al. 1999b]{rem99b}Remillard, R. A., Morgan, E. H., McClintock, Bailyn, C. D., \& Orosz, J. A. 1999b, \apj, 522, 397

\bibitem[Remillard et al. 2002]{rem02}Remillard, R. A., Sobczak, G. J., Muno, M., \& McClintock, J. E. 2002, \apj, 564, 962

\bibitem[Revnivtsev, Gilfanov, \& Churazov 2000]{rev00}Revnivtsev, M., Gilfanov, M., \& Churazov, E. 2000, A\&A, 363, 1013

\bibitem[Shahbaz et al. 1999]{sha99}Shahbaz, T., et al. 1999, \mnras, 306, 89

\bibitem[Shapiro \& Teukolsky 1983]{sha83}Shapiro, S. L., \& Teukolsky, S. A. 1983 $Black ~Holes, White ~Dwarfs, ~and ~Neutron ~Stars$, (New York: Wiley)

\bibitem[Silbergleit, Wagoner, \& Ortega-Rodriguez 2001]{sil01}Silbergleit, A. S., Wagoner, R. V., \& Ortega-Rodriguez, M. 2001, \apj, 548, 335

\bibitem[Strohmayer 2001a]{str01a}Strohmayer, T. E. 2001, \apjl, 552, L49

\bibitem[Strohmayer 2001b]{str01b}Strohmayer, T. E. 2001, \apjl, 554, L169

\bibitem[Sobczak et al. 1999]{sob99}Sobczak, G. J., McClintock,
J. E., Remillard, R. A., Bailyn, C. D. \& Orosz, J. A. 1999, \apj, 520, 776

\bibitem[Sobczak et al. 2000a]{sob00a}Sobczak, G. J., McClintock,
J. E., Remillard, R. A., Cui, W., Levine, A.  M., Morgan, E. H.,
Orosz, J. A., \& Bailyn, C. D. 2000a, \apj, 531, 537

\bibitem[Sobczak et al. 2000b]{sob00b}Sobczak, G. J., McClintock,
J. E., Remillard, R. A., Cui, W., Levine, A.  M., Morgan, E. H.,
Orosz, J. A., \& Bailyn, C. D. 2000b, \apj, 544, 993

\bibitem[Swank, Smith, \& Markwardt 2002]{swa02}Swank, J., Smith, E., \& Markwardt, C. 2002, IAU Circ. 7792

\bibitem[Tomsick et al. 2001a]{tom01a}Tomsick, J. A., Smith, D. A., Swank, J. H., Wijnands, R., \& Homan, J. 2001, IAU Circ. 7575

\bibitem[Tomsick et al. 2001b]{tom01b}Tomsick, J. A., Corbel, S., \& Kaaret, P. 2001, \apj, 563, 229

\bibitem[Wagoner 1999]{wag99}Wagoner, R. V. 1999, Physics Reports, 311, 259

\bibitem[Wagoner, Silbergleit, \& Ortega-Rodriguez 2001]{wag01}Wagoner, R. V., Silbergleit, A. S., \& Ortega-Rodriguez, M. 2001, \apj, 559, L25

\bibitem[Wijnands et al. 1999]{wij99a}Wijnands, R., Homan, J., \& van
der Klis, M. 1999, \apj, 526, 33

\bibitem[Zhang, Cui, \& Chen 1997]{zha97}Zhang, S. N., Cui, W., \& Chen, W. 1997, \apj, 482, L155

\end{thebibliography}
\end{document}